\documentclass[11pt,letterpaper]{article}

\usepackage{jheppub}
\usepackage{color}
\usepackage{graphicx}
\usepackage{wrapfig,enumerate,slashed}

\usepackage{footmisc}
\usepackage{amsmath}
\usepackage{wasysym} 
\usepackage{graphicx}
\usepackage{color}
\usepackage{comment}
\usepackage{hyperref}

\usepackage{amssymb}
\usepackage{amsmath}
\usepackage{epstopdf}
\usepackage{amsmath}
\usepackage{multirow}

\usepackage{url}
\usepackage{color}
\usepackage{listings}

\newcommand{\bee}{\begin{eqnarray}}
\newcommand{\eee}{\end{eqnarray}}
\newcommand{\beeq}{\begin{equation}}
\newcommand{\eeeq}{\end{equation}}

\newcommand{\fb}{{\text{fb}}}

\newcommand{\be}{\begin{equation}} \newcommand{\ee}{\end{equation}}
\newcommand{\ba}{\begin{eqnarray}} \newcommand{\ea}{\end{eqnarray}}
\newcommand{\nn}{\nonumber} \renewcommand{\bf}{\textbf}

\newcommand{\GeV}{\text{~GeV}}
\newcommand{\TeV}{\text{~TeV}}

\definecolor{red1}{cmyk}{0,1,1,0.1}
\definecolor{blue1}{cmyk}{1,0,0,0}

\begin{document}

\title{Cornering diphoton resonance models at the LHC}

\author[a]{Mihailo Backovi\'{c},}
\author[b]{Suchita Kulkarni,}
\author[c]{Alberto Mariotti,}
\author[d]{Enrico Maria Sessolo}
\author[e]{and Michael Spannowsky}

\affiliation[a]{Center for Cosmology, Particle Physics and Phenomenology (CP3), Universite Catholique de Louvain,\\ 
B-1348 Louvain-la-Neuve, Belgium}
\affiliation[b]{Institute of High Energy Physics, Austrian Academy of Sciences,\\ 
Nikolsdorfergasse 18, 1050 Vienna, Austria}
\affiliation[c]{Theoretische Natuurkunde and IIHE/ELEM, Vrije Universiteit Brussel, and International Solvay Institutes, \\
Pleinlaan 2, B-1050 Brussels, Belgium}
\affiliation[d]{National Centre for Nuclear Research, \\
Ho{\. z}a 69, 00-681 Warsaw, Poland}
\affiliation[e]{Institute for Particle Physics Phenomenology, Department of Physics, Durham University, \\
DH13LE, UK}

\emailAdd{mihailo.backovic@uclouvain.be}
\emailAdd{suchita.kulkarni@oeaw.ac.at}
\emailAdd{alberto.mariotti@vub.ac.be}
\emailAdd{enrico.sessolo@ncbj.gov.pl}
\emailAdd{michael.spannowsky@durham.ac.uk}

\abstract{
We explore the ability of the high luminosity LHC to test models which can explain the 750 GeV diphoton excess. 
We focus on a wide class of models where a 750 GeV  singlet scalar couples to Standard Model gauge bosons and quarks, as well as dark matter.
Including both gluon and photon fusion production mechanisms, 
we show that LHC searches in channels correlated with the diphoton signal will be able to probe wide classes of diphoton models with
$\mathcal{L} \sim 3000\, \fb^{-1}$ of data. Furthermore, models in which the scalar is a portal to the dark sector can be cornered with as little as $\mathcal{L} \sim 30\, \fb^{-1}$.}

\preprint{IPPP/16/46, DCPT/16/92,CP3-16-23, HEPHY-PUB 967/16}

\maketitle


\section{Introduction}
\label{sec:intro}

Both ATLAS and CMS collaborations recently announced an excess in the diphoton spectrum around the invariant mass of $m_{\gamma \gamma} \approx 750 \GeV$. While the excess is not statistically significant to claim a discovery (ATLAS finds a local significance of $3.6 \sigma$~\cite{ATLAS-CONF-2015-081,ATLAS-CONF-2016-018} and CMS one of $3.0 \sigma$~\cite{CMS-PAS-EXO-15-004,CMS-PAS-EXO-16-018}), it is certainly interesting to entertain the idea that the data points to the existence of a new particle. 

If such particle is a singlet under the SM gauge group, it is inevitable that the diphoton excess will be correlated with signals in other channels involving gauge bosons (e.g. $Z\gamma$, $ZZ$, or $WW$). It has been shown that an excess should appear at least in one of the before-mentioned channels, regardless of the underlying model parameters~\cite{Low:2015qho,Kamenik:2016tuv}. In the optimistic scenario where the 750 GeV diphoton excess remains as more data comes in, measurements of other final states which are correlated to the diphoton excess will hence become instrumental in both confirming the signal, as well as determining the properties of the new particle. In particular, not observing correlated signals in final states with Standard Model (SM) gauge bosons will have direct implications on many  scenarios attempting to explain the excess.

The width of the diphoton excess offers additional crucial information about the nature of the possible new particle. The line-shape of the excess measured by ATLAS indicates a rather broad resonance with a width $\Gamma_{\rm tot} \simeq 45$~GeV, which is difficult to account for if it decays only into Standard Model (SM) particles.  Large unobserved decay modes can point to interactions between the new resonance and dark matter, leading to collider signatures in channels with large missing energy, as well as signals in direct dark matter detection experiments via scattering off nuclei, and the measurements of galactic $\gamma$-ray fluxes~\cite{Backovic:2015fnp,Barducci:2015gtd, Mambrini:2015wyu,D'Eramo:2016mgv}.

In this paper we explore the reach of LHC run-2 searches for the diphoton resonance models. 
Leading order approximations for the production of $s$-channel resonances allow for the use of simple scaling rules to study the constraints of existing and future LHC results on the model parameter space. 
Specifying the features for the diphoton excess, such as the production mechanism and cross section, defines a hyper-surface in the multi-dimensional parameter space which can explain the excess.
Our approach consists in constraining these surfaces further, by imposing collider bounds on correlated final states.

Using concrete examples, we demonstrate the most sensitive channels and relevant bounds, as well as the required integrated luminosity to rule out particular models explaining the diphoton excess. For concreteness, we assume throughout the paper that the resonance is a scalar singlet under the SM gauge group. Hence, its interactions with SM particles are captured at leading order by a set of dimension-5 operators suppressed by a new physics scale $\Lambda$~\cite{Franceschini:2015kwy}. We further assume that the new resonance does not mix with the SM Higgs boson, as existing and projected limits from Higgs coupling measurements set strong indirect constraints \cite{Falkowski:2015swt}.

We discuss three concrete benchmark scenarios, which serve to encompass a large class of 750 GeV diphoton resonance models.  
First, we study the ``vanilla'' scenario, in which a scalar singlet couples only to SM gauge bosons via dimension-5 effective interactions. 
Second, we consider a scenario in which decays of a 750 GeV scalar into an invisible sector ($i.e.$ dark matter)  
accommodate the potentially large resonance width. Finally, we analyze a scenario in which the scalar is allowed to couple to SM quarks in addition to SM gauge bosons. 
For the purpose of studying future LHC limits on the three scenarios, we project existing 8 and 13~TeV limits on production of gauge boson, mono-jet, and $t\bar{t}$ final states at various 
luminosities. 
We outline the strategy we adopt and the simplified approach we employ to project limits for the LHC 
in Sec.~\ref{sec:setup}. In Sec.~\ref{sec:models} we present our main results, where we confront concrete diphoton scenarios with the existing LHC bounds and our 
estimated projections for the 13~TeV run. Finally, we briefly summarize our results and conclude in Sec.~\ref{sec:sum}. In Appendix~\ref{sec:limit} we provide more technical details about limit projection and in Appendix~\ref{sec:anfor} we 
review the analytical forms used here for the calculation of the decay widths.



\section{General strategy and LHC limits}
\label{sec:setup}

We begin with a brief discussion of the possible production modes for the 750 GeV diphoton resonance. 
We limit our discussion to the case of a pure scalar, however most of the qualitative conclusions in our paper will hold in the case of a pseudo-scalar resonance as well. 
In the most general scenario, the onshell production cross section of the scalar resonance can be approximated by
$$
	\sigma (pp \rightarrow S) \approx \sum_{ij} C_{ij} (s, M) \sigma_{ij}, 
$$
where $i, j$ are proton constituents (including photons), $C_{ij}$ are the dimensionless parton luminosity factors and $\sigma_{ij}$ are partonic cross sections. Limits from 8 TeV LHC disfavor production via light quarks \cite{Franceschini:2015kwy, Gupta:2015zzs} and we will hence limit ourselves to scenarios in which the new scalar particle is produced via either gluon fusion ($gg$) or photon fusion ($\gamma \gamma$) initial state. 

The photon fusion production mechanism deserves further discussion. 
Production of a scalar resonance compatible with the diphoton excess via photon fusion are studied in Refs.~\cite{Csaki:2015vek,Csaki:2016raa,Harland-Lang:2016qjy,Abel:2016pyc,Fichet:2015vvy}.\footnote{First coupling 
constraints for such models using 8 TeV data have been obtained in~\cite{Jaeckel:2012yz}.}
However, it is important to note that many subtleties arise in considering the photon fusion channel. The cross section enhancement between 8 and 13 TeV center-of-mass energy at the LHC is subject to large uncertainties and can vary between a factor 2 and 4~\cite{Csaki:2016raa,Harland-Lang:2016qjy}. Hence, pure photon production is possibly already in tension with 8 TeV data if the ratio is closer to 2. In addition, given the inclusive nature of the diphoton excess measurements in the ATLAS and CMS searches, it is also possible that vector boson fusion (VBF) channels with one or two additional reconstructed jets contribute to the overall production cross section.  
We estimated the VFB contributions with one or two additional jets for the models we consider in this paper. We found that VBF contributes at most $\sim 15\%$
 of the inclusive diphoton production cross section in the regions of the parameter space compatible with the observed diphoton excess.\footnote{The full treatment of multi-jet merging in electroweak processes is beyond the scope of our paper~\cite{Alwall:2007fs}.}   
We will thus neglect such VBF contributions in the following.

Continuing, within the narrow width approximation the diphoton cross section at leading order is simply
\begin{eqnarray}
	\sigma_{\gamma \gamma} &=& \left[ \sigma_{\gamma} (pp \rightarrow S) + \sigma_{g} (pp \rightarrow S) \right] \times \textrm{Br} (S \rightarrow \gamma \gamma) \nonumber \\
						&=&  \left[ c_\gamma^2 \sigma_{\gamma} (pp \rightarrow S) _{c_\gamma = 1} + c_G^2 \sigma_{g} (pp \rightarrow S) _{c_G = 1} \right] \times \textrm{Br} (S \rightarrow \gamma \gamma) \,,
	\label{eq:sigma} 
\end{eqnarray}
where we have factored out the dependence on $S$ couplings to gluons and photons ($c_G$ and $c_{\gamma}$). $\sigma_{\gamma, g}$ are the photon and gluon initiated production cross sections respectively. 
Note that $\textrm{Br}(S\rightarrow \gamma \gamma)$ 
is an implicit function of all of the theory parameters.
Assuming a signal cross section $\sigma^{*}_{\gamma \gamma}$, consistent with the observed excess, Eq.~\eqref{eq:sigma} can be solved for $c_G^*$ as a function of the remaining parameters in a given model, hence defining a slice of the parameter space which can accommodate the excess. 
Note that in the limit of $c_\gamma \rightarrow 0$ the branching ratio into photons also vanishes, yielding no viable solution for $c_G^*$.

Parameter space slices determined by $\sigma_{\gamma \gamma}^*$ can then be bound by searches in the complementary final state channels.  ATLAS and CMS have recently published the first results from 
the LHC 13~TeV run, with an integrated luminosity of $3.2\textrm{ fb}^{-1}$  and $2.3 \textrm{ fb}^{-1}$ respectively, which can be used to constrain existing models. Bounds from resonance searches involving gauge bosons final states are of particular relevance for constraining gauge invariant parameterizations of the diphoton models. 

\begin{table}
\footnotesize\begin{center}
\begin{tabular}{ccccccc}
\multirow{2}{*}{Search}& 8 TeV limit [fb]& 13 TeV limit [fb] & \multicolumn{4}{c}{13 TeV limit [fb] (expected)}  \\
&(observed)& (observed) & $\mathcal{L} = 3.2\, \fb^{-1}$  & $\mathcal{L} = 30\, \fb^{-1}$  & $\mathcal{L} = 300\, \fb^{-1}$  &$\mathcal{L} = 3000\, \fb^{-1}$ \\
\hline 
$Z\gamma$&11~\cite{Aad:2014fha} & 30~\cite{ATLAS-CONF-2016-010}  & 43 & 14 & 4.4 & 1.4\\
$ZZ$&12~\cite{Aad:2015kna} & 180~\cite{ATLAS-CONF-2015-071} & 82 & 27 & 8.5 & 2.7 \\
$WW$ & 40~\cite{Aad:2015agg} & 400~\cite{ATLAS-CONF-2016-021} & 300 & 98 & 31 & 9.8 \\
$t\bar{t}$& 460~\cite{Chatrchyan:2013lca} & 10000~\cite{ATLAS-CONF-2016-014} & 3267 & 1067 & 337 & 107 \\
MET+$j$& 7.2 (SR7)~\cite{Aad:2015zva} & 61 (IM5)~\cite{Aaboud:2016tnv} & 51  &  &  &  \\
 &  & 19 (IM7)~\cite{Aaboud:2016tnv} & 15 & 5 & 1.5 & 0.5 \\
\hline

\end{tabular}
\caption{Extrapolations of experimental limits relevant for the 750 GeV diphoton. 
The models are constrained by the strongest of the 8~TeV and 13~TeV observed limits. 
The inclusive regions SR7 (for the mono-jet (MET+$j$) 8~TeV search) and IM5 (for the corresponding 13~TeV results) are charecterized by $E_T^{\textrm{miss}}>500$~GeV. The inclusive region IM7 for the 13~TeV search is defined by  $E_T^{\textrm{miss}}>700$~GeV. For the $ZZ$ and $t\bar{t}$ searches the expected limit at $3.2\textrm{ fb}^{-1}$ is extrapolated from the  8~TeV expected bound (see text).} 
\label{tab:extrapolCS}
\end{center}
\end{table} 

We present a summary of the bounds used in this paper in Table~\ref{tab:extrapolCS}. In the $Z\gamma$ final state the 95\%~C.L. 8~TeV ATLAS upper bound on the production cross section times branching ratio~\cite{Aad:2014fha} reads approximately 11~fb, whereas the bound from the equivalent search in Run~2~\cite{ATLAS-CONF-2016-010} yields $\sim 30\textrm{ fb}$. While the data from Run~2 is not particularly useful to constrain these scenarios yet, it can nonetheless be used to estimate the reach of these searches for future luminosity. The idea is based on the assumption that, while being model dependent, quantities like cross sections, acceptances and efficiencies do not depend on the integrated luminosity.
In the limit of a large number of events,  one can obtain the expected 95\%~C.L. cross section bound at any target luminosity $\mathcal{L}$ by 
rescaling the $3.2\textrm{ fb}$ limit with the ratio of corresponding luminosities.  
Considering, for example, the $Z\gamma$ case in Table~\ref{tab:extrapolCS}, rescaling the expected $3.2\textrm{ fb}^{-1}$ bound of 
$\sim 43\textrm{ fb}$~\cite{ATLAS-CONF-2016-010} yields the projected values shown in the columns of 30, 300, and $3000\textrm{ fb}^{-1}$.\footnote{We stress that the limits we obtain in this way are conservative. Data-driven methods can reduce systematic 
uncertainties when large data samples are available and dedicated reconstruction techniques~\cite{Abdesselam:2010pt,Altheimer:2012mn}. 
exploiting the increased center-of-mass energy at $13/14$ TeV and different decay mode
scan improve on the limits we extrapolate.}

We use the above luminosity-rescaling ansatz to obtain the majority of the projections considered in this paper. However, while luminosity rescaling provides conservative estimates in most cases, it does not always reproduce the most realistic expectations. As experience with the large number of search results produced during and after the 8~TeV run has shown, a statistical combination of the data obtained in searches sensitive to different final states often leads to a dramatic improvement in the bounds with respect to searches in single channels. 
For instance, a direct comparison of the expected 8~TeV bounds on the production cross section of a heavy scalar decaying to $ZZ$ in the $llll$, $ll(\nu\nu)qq$, $ll\nu\nu$, and a combination thereof~\cite{Aad:2015kna} shows that the combined limit is at least a factor of two stronger than any of the individual bounds.  ATLAS has published results for the 13~TeV $ZZ$ resonance searches in the $\nu \nu qq$~\cite{ATLAS-CONF-2015-068} and $ll qq$~\cite{ATLAS-CONF-2015-071} final states, but at this early stage  the combination has not been published. It is reasonable to assume that the final combined limit will be also stronger than the one obtained in Refs.~\cite{ATLAS-CONF-2015-068} or~\cite{ATLAS-CONF-2015-071}. Hence, we will adopt the 13 TeV $ZZ$ limit extrapolated from the combined 8 TeV LHC limit, using the procedure described in detail in Appendix~\ref{sec:limit}. We have verified that the procedure accurately reproduces the existing 13 TeV limits in the $ll qq$ and $\nu \nu qq$ channels, leading us to conclude that our combined limit extrapolation is also accurate (see Appendix~\ref{sec:limit} for more details).

Limits on the resonant $WW$ production both at 8 TeV and 13 TeV exist \cite{Aad:2015agg, ATLAS-CONF-2016-021}, 
and we adopt the observed limits on the 750 GeV resonance from both LHC runs. 

The strongest observed ATLAS limits in the final state with at least one jet and large missing transverse momentum $E_T^{\textrm{miss}}$ 
(hereafter MET+$j$) 
comes from inclusive search bins denominated SR7 (in the 8~TeV search~\cite{Aad:2015zva}) and IM5 (at 13~TeV \cite{Aaboud:2016tnv}), which are defined by $E_T^{\textrm{miss}}>500 \GeV$. The strongest expected limit at 13~TeV comes instead from the inclusive bin IM7 with $E_T^{\textrm{miss}}>700 \GeV$. Hence, for the purpose of extrapolating the limit to higher luminosities
we use the expected limit at 13~TeV in the inclusive bin IM7.

Finally, current experimental searches for $t\bar{t}$ resonances at 13~TeV~\cite{ATLAS-CONF-2016-014} 
have focused only on the boosted regime, with no publicly available result on searches for $t\bar{t}$ resonances in the resolved regime. Boosted top analyses are ill suited for efficient reconstruction of the $t\bar{t}$ final states with invariant mass of $\lesssim 1 \text{TeV}$ (assuming the standard fat jet cone of radius $R=1.0$), resulting in 13~TeV limits on a 750~GeV resonance which are far weaker than the extrapolated 8~TeV limits in the resolved jet analysis. For 13~TeV $t\bar{t}$ final state, we hence adopt an extrapolated limit from the resolved 8~TeV analysis, obtained with the techniques explained in Appendix \ref{sec:limit}.



\section{Diphoton resonance models}
\label{sec:models}

In order to illustrate the strategy we have discussed in the previous section, 
we consider a concrete set of models where the new resonance is represented by a singlet scalar coupled to the SM with dimension-five operators.
Moreover, we also investigate the possibility that the new resonance plays the role of a portal 
to a dark sector. 
A wide class of diphoton resonance models can comprehensively be described by the interaction Lagrangian
\be
\label{Lag_full}
	\mathcal{L} 
	\supset 
	\frac{c_{G}}{\Lambda}S G^{\mu\nu} G_{\mu\nu}+\frac{c_{W}}{\Lambda}S W^{\mu\nu} W_{\mu\nu}+\frac{c_{B}}{\Lambda}S B^{\mu\nu} B_{\mu\nu}
	+ g_{f} \sum_{q} \frac{m_q}{\Lambda} S \bar q q + g_{X} S \bar X X\,,  
\ee
where $G_{\mu\nu}$, $W_{\mu\nu}$, and $B_{\mu\nu}$ are the $SU(3)$, $SU(2)$, and $U(1)$ field strength tensors, respectively, $q$ indicates SM fermions (of mass $m_{q}$), and $X$ is an invisible Dirac fermion which can play the role of dark matter.
In the following we will independently study different subsets of this general class of models by switching on and off some of the couplings in Eq.~(\ref{Lag_full}).

Note that we assumed that the new scalar resonance does not couple to the SM Higgs boson. The coupling to the Higgs is mainly constrained by the allowed size of the mixing angle, which is bounded by LHC Higgs coupling measurements to be $\lesssim O(10-20\%)$ \cite{Falkowski:2015swt}. 
This already puts significant constraints on possible correlated signals of the new resonance in the Higgs final states, and we leave to future studies a detailed investigation of the LHC 13~TeV reach for these signatures.

We point out that the couplings of the scalar are chosen proportional to the quark masses, to respect minimal flavor violation. Since $S$ is a singlet of the SM gauge groups, the new couplings to SM fermions should be considered as descending from dimension-five 
operators such as $\frac{1}{\Lambda} y S H \bar Q_L u_R$, which after electroweak symmetry breaking, generate the couplings in Eq.~\eqref{Lag_full}. The couplings with SM fermions in Eq.~\eqref{Lag_full} have an extra suppression factor scaling, $m_q/\Lambda$, for this reason.
Without loss of generality, we have introduced a unique suppression scale $\Lambda$ for the various operators, which are then weighted by different $O(1)$ couplings $( c_{G},c_{W},c_{B},g_f )$. For definiteness we will take $\Lambda=10~\text{TeV}$ throughout the paper. 

As mentioned before, we will consider a combination of production mechanisms. For couplings of similar size gluon-fusion is typically the dominant production mechanism.  However we will explore also regions of 
the parameter space where photon-fusion processes, which scale like $c_{\gamma}^2 \equiv (c_{B} \cos^2\theta_W +c_{W} \sin^2\theta_W)^2 $, are dominating.
In the case of quark-initiated production, the dimensionless Yukawa couplings of $S$ to the quarks are 
suppressed by a factor $g_f m_q/\Lambda$, as they descend from higher dimensional gauge invariant operators. Hence, the light quark contributions are suppressed by the small quark masses, while the heavy quark ones are suppressed by small proton PDF and by the smallness of  $g_f m_q/\Lambda$  (since we are considering $O(1)$ couplings and $\Lambda \gg m_q$).  In particular, the top loop induced gluon fusion contribution to the $S$ production cross section
is negligible with respect to other production mechanisms in the range of couplings that we study.

In order to estimate the production cross section for the resonance $S$ through the available processes we make use of several tools. We have implemented the model of Eq.~\eqref{Lag_full} in FeynRules~\cite{Alloul:2013bka} and we simulate the production of $S$ at the LHC using MadGraph5\_aMC@NLO (MG5\_aMC)~\cite{Alwall:2014hca} with the NN23LO1~\cite{Ball:2012cx}  PDF set for gluon as well as for photon PDFs. For photon-fusion we consider both the inelastic-inelastic as well as the elastic-inelastic proton scattering processes.

Given the production cross section for the resonance, the cross sections in the various final states are determined by the branching ratios. Analytic formulas for the partial decay widths of $S$ in the model \eqref{Lag_full} are listed in Appendix~\ref{sec:ana_formulae}.

In exploring the parameter space of the model, our strategy relies on solving the condition $\sigma_{\gamma\gamma} = \sigma_{\gamma\gamma}^*$ (see Eq.~\eqref{eq:sigma}) for the coupling $c_{G}$. After fixing the couplings $g_{X}$ and $g_f$ to some representative value, we present the results in the $(c_{B},c_{W})$ plane. 
For definiteness we choose $\sigma_{\gamma\gamma}^* =7\textrm{ fb}$ but our results are qualitatively 
robust under change of the required cross section.
We will also display the $c_{G}$ contours necessary to fit the excess, 
and identify the most relevant production mechanism on each region of the parameter space.

\subsection{The ``vanilla'' model: $g_{X}=g_{f}=0$}\label{sec:vanilla}

We start our analysis by considering the simplest version of the model capable of explaining the diphoton excess, $i.e.$ we set the couplings to dark matter and SM fermions to 0. The so called ``vanilla'' model is then parameterized only by three couplings: $c_W, c_B$ and $c_G$. We explore the parameter space in the range $(c_{B},c_{W}) \in \{-1,1\}$ and for every value of $(c_{B},c_{W})$ we solve the equation $\sigma_{\gamma \gamma}=\sigma_{\gamma \gamma}^* = 7$~fb for $c_{G}^2$, imposing the conservative bound $c_{G} < 4 \pi$.

\begin{figure}[!htb]
\begin{center}
\includegraphics[width=0.32\textwidth]{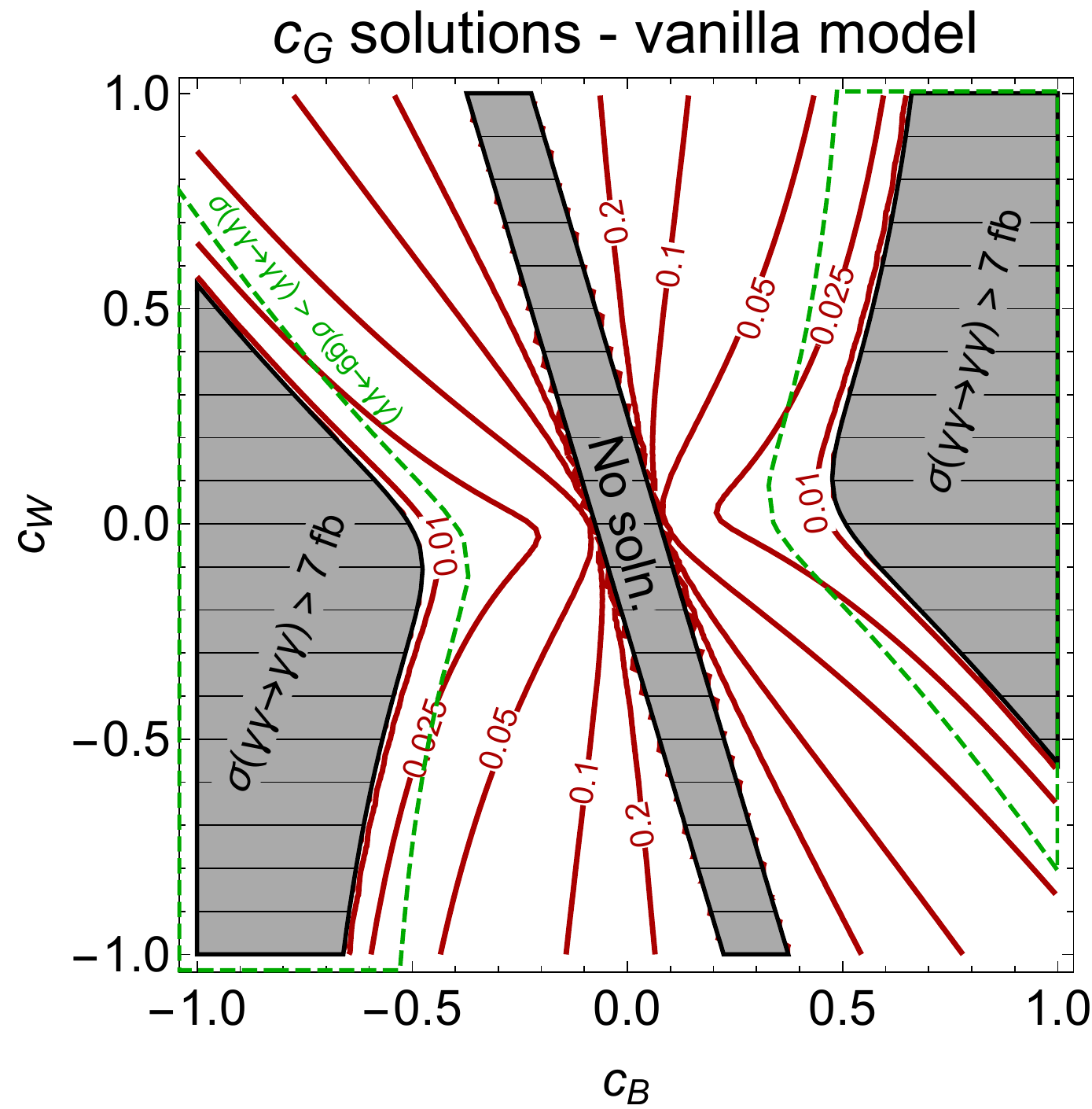}
\includegraphics[width=0.32\textwidth]{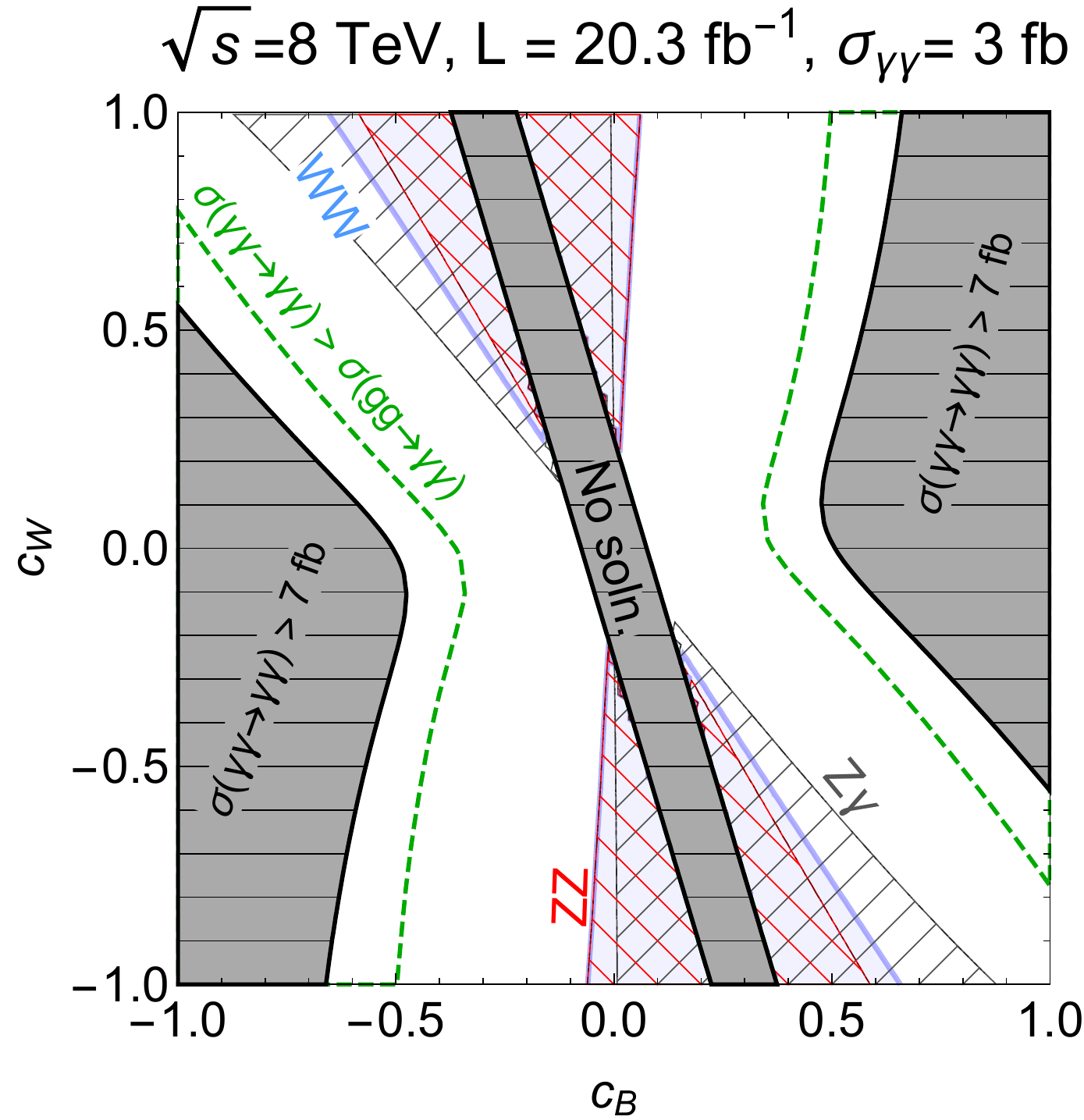} 
\includegraphics[width=0.32\textwidth]{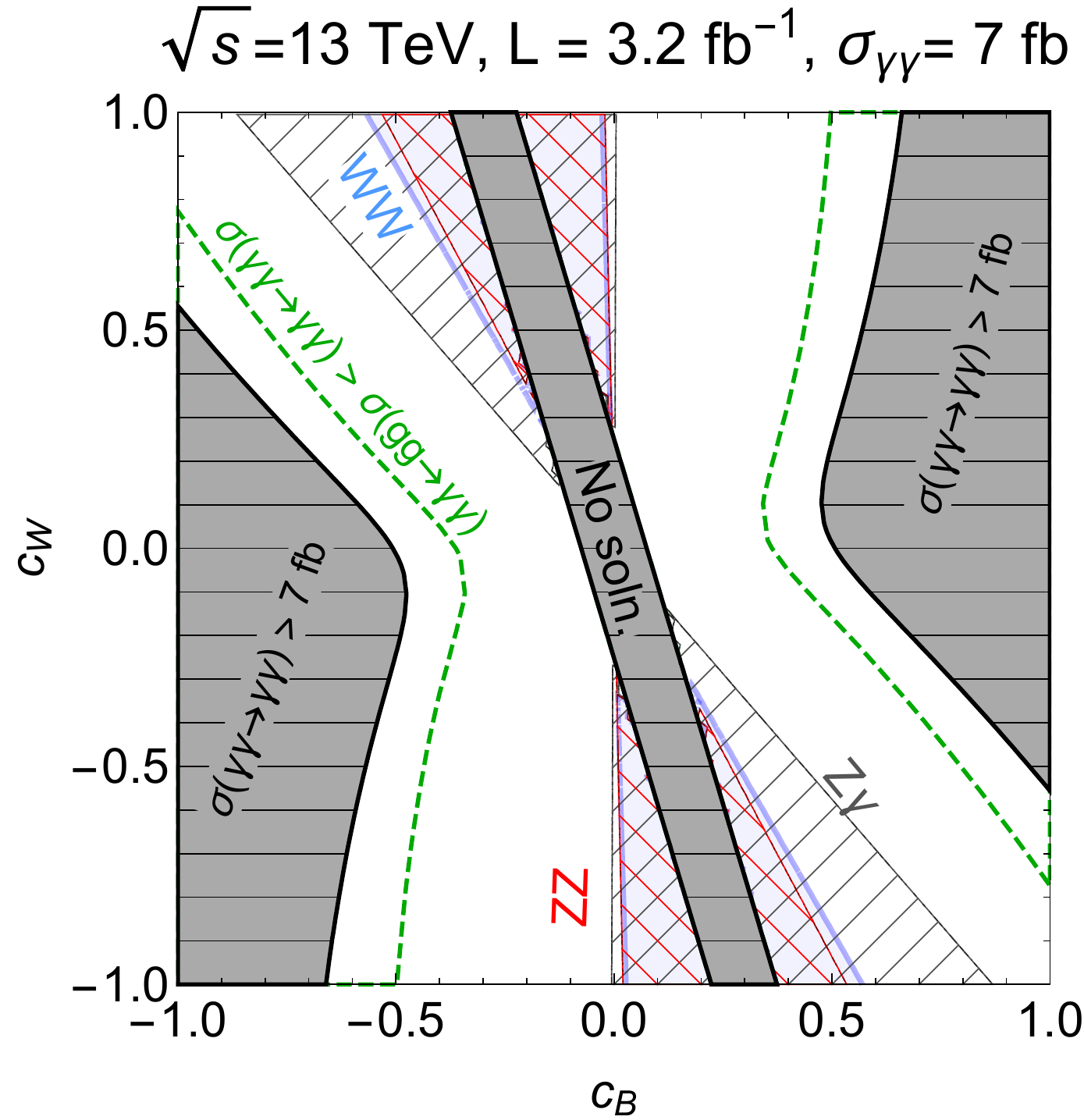} \\
\includegraphics[width=0.32\textwidth]{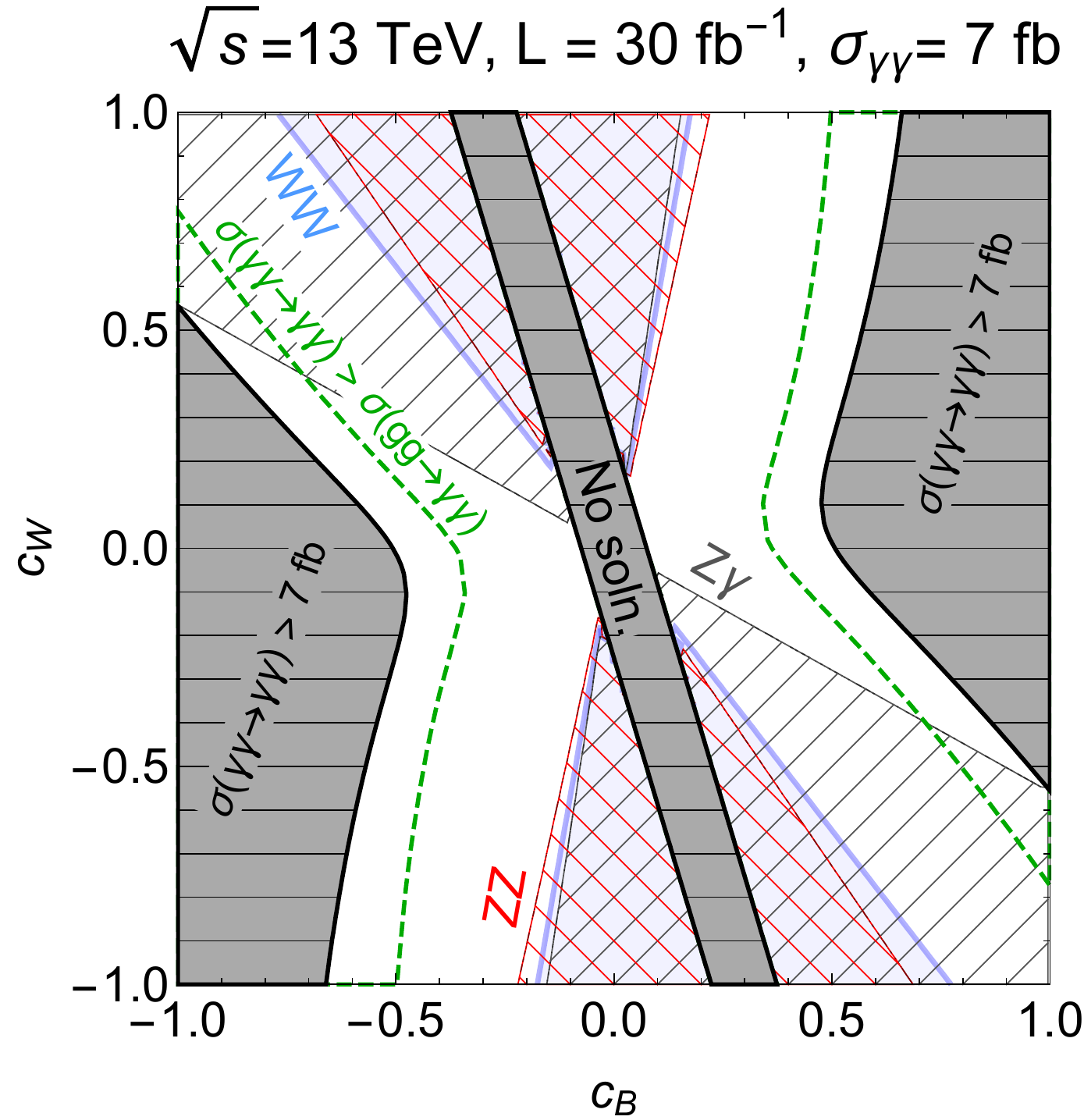} 
\includegraphics[width=0.32\textwidth]{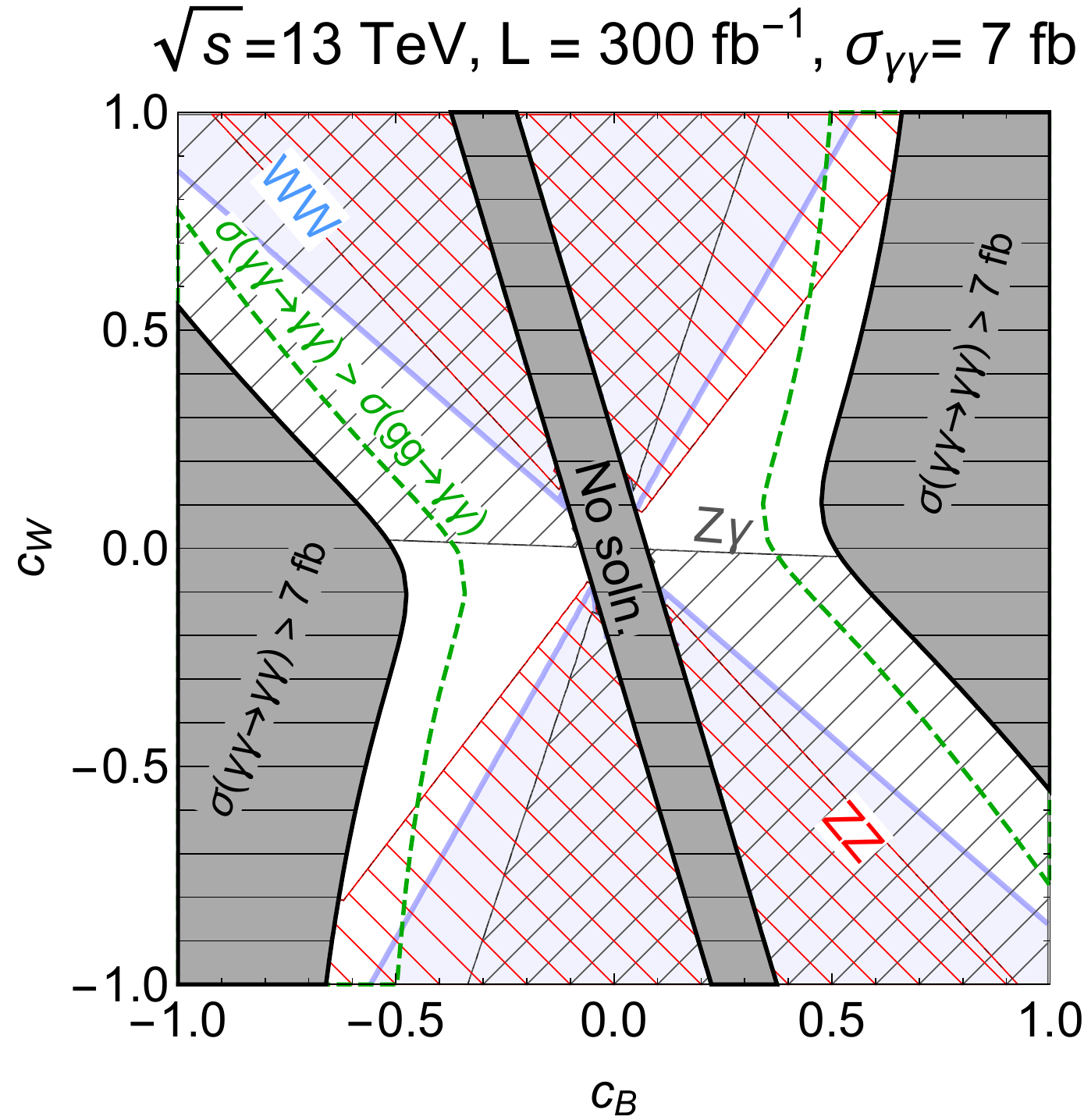}
\includegraphics[width=0.32\textwidth]{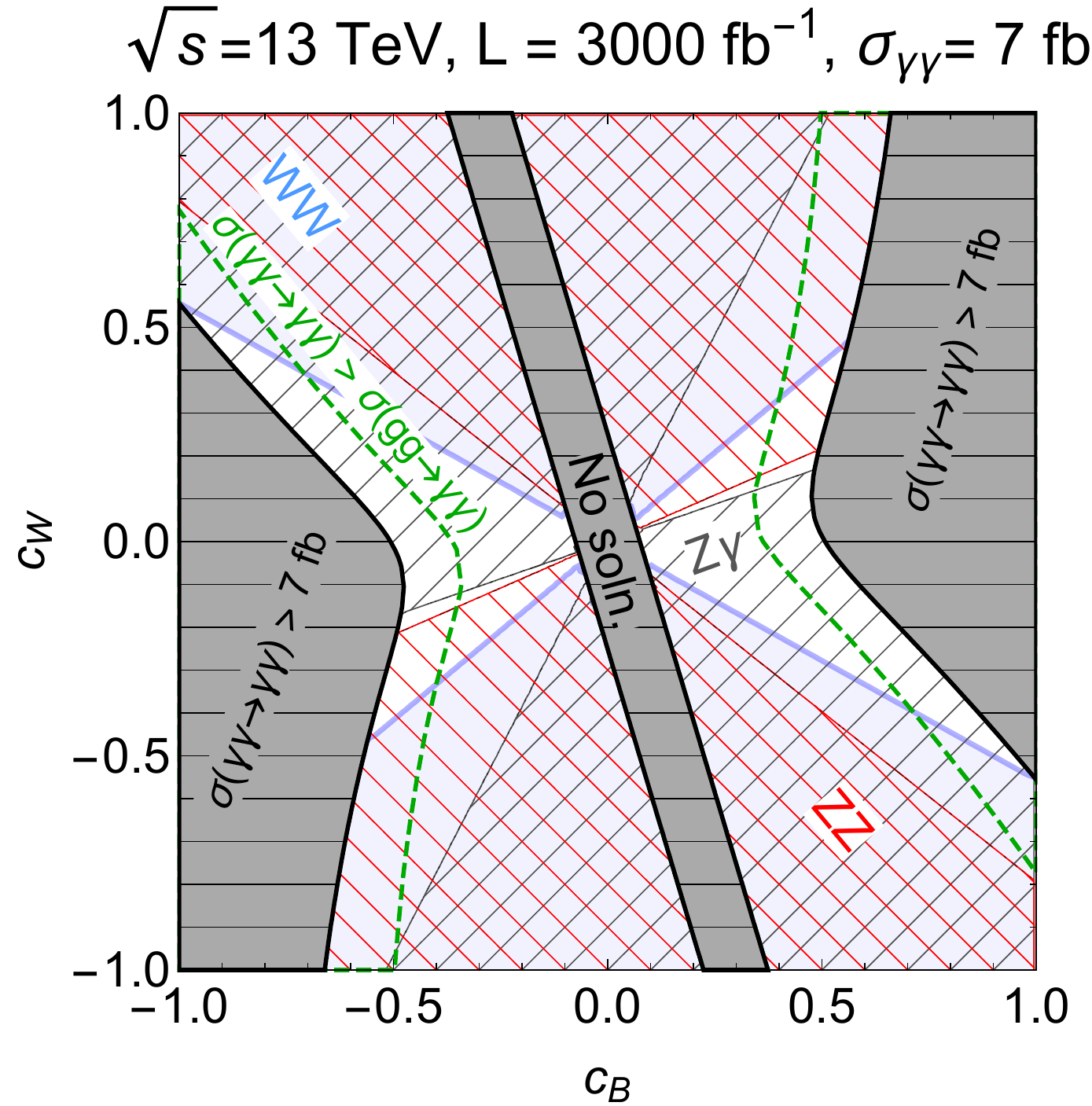}
\caption{Analysis of the ($c_B$, $c_W$) plane in the vanilla model parameter space. We apply the bounds from the 8 and 13~TeV runs of the LHC as well as present several projections in future luminosities.
The dark grey shaded regions are where there we find no $c_G$ solution for the diphoton excess. Other shaded regions are excluded by particular searches labeled on the plots. The upper left panel shows the contours of constant $c_G$, necessary to accommodate the required signal cross section.}
\label{fig:vanillamodel} 
\end{center}
\end{figure}

Figure~\ref{fig:vanillamodel} shows our first result. In the upper left plot of Fig.~\ref{fig:vanillamodel} we display in solid red the contours of $c_G$ consistent with signal cross section $\sigma_{\gamma \gamma}^*$. 
The values of $c_G$ decrease towards larger values of $c_B$ and $c_W$ since the branching ratio into photons increases. In addition, the photon fusion contribution to the total production cross section also increases with larger $c_B$ and $c_W$  values, requiring a lower gluon fusion contribution to reproduce the signal. 

The green dashed contours mark the boundary between regions where gluon fusion dominates and regions where photon fusion dominates  instead.
Photon fusion can be dominant only for large values of $c_B$. The shape of the green dashed contour is determined by the competition between the BR($S \to \gamma \gamma$) and the branching ratios for the other electroweak bosons ($Z\gamma$, $ZZ$, and $WW$), 
which can deplete the signal in $\gamma \gamma$.

The gray regions indicate regions where no solutions for $c_G$ resulting into $\sigma_{\gamma \gamma}=\sigma_{\gamma \gamma}^*$ exist. We can identify two distinct gray areas which have different physical interpretations.
The almost vertical gray stripe close to the central axis (denoted ``No-soln.'') 
is located around the straight line $c_{\gamma}=0$.  In this regime, the coupling to
photons is very small, leading to the fact that no real value of $c_G$ can reproduce the signal strength $\sigma_{\gamma\gamma}^*$.  The argument can be understood analytically as follows. The coupling to photons is almost vanishing in the central grey region,  leading to a gluon fusion dominated production mechanism. We can then write
\be
\sigma_{gg} ( p p \to S \to \gamma \gamma) = \frac{\mathcal{C}_{gg}}{m_{S} s} \frac{\Gamma_{gg} \Gamma_{\gamma \gamma}}{\Gamma_{\mathrm{tot}}}\,,
\ee
where $\mathcal{C}_{gg}$ is the gluon luminosity and $s$ is the centre of mass energy.
One can impose $\sigma_{gg}( p p \to S \to \gamma \gamma)=\sigma_{\gamma \gamma}^*$ and solve this equation for the total width of $S$ obtaining 
\be
\Gamma_{\mathrm{tot}} = \frac{\mathcal{C}_{gg}}{ \sigma_{\gamma \gamma}^* m_{S} s} \Gamma_{gg} \Gamma_{\gamma \gamma}\,.
\ee
Given that the total width of the resonance is always larger or equal than the width into gluons, $\Gamma_{\rm tot} \geq \Gamma_{gg}$, we arrive to the
inequality
\be
\Gamma_{\gamma\gamma} \geq \frac{ \sigma_{\gamma \gamma}^* m_{S} s}{\mathcal{C}_{gg}}\,,
\ee
which implies an absolute lower bound for the partial decay width into photons, necessary to accommodate the diphoton excess.
Inserting the explicit expression for the $\gamma \gamma$ partial width (see Appendix \ref{sec:ana_formulae}) we obtain the lines which delimit the vertical gray stripe:
\be
c_{W} = -c_{B} \tan^{-2} \theta_W \pm \frac{2 \sqrt{\pi s \sigma^*} \Lambda}{ m_S \sqrt{\mathcal{C}_{gg}}\sin^2 \theta_W}\,.
\ee
The above argument does not depend on the other contributions to $\Gamma_{\rm tot}$ and is therefore a generic result for the complete model of Eq.~\eqref{Lag_full}, independently of the value of $g_{X}$ and $g_f$. We will indeed find the same gray stripe around $c_{\gamma}=0$ in all of the other scenarios considered in this paper.

The other gray region, denoted with $\sigma_{\gamma\gamma} > 7$~fb in Fig.~\ref{fig:vanillamodel}, are instead characterized by excessively large rates in $\gamma \gamma$, completely dominated by photon-fusion processes. The internal border of the region identifies the line where the production mechanism is $100\%$ photon fusion, and
$c_{G}=0$. 

\begin{figure}[!]
\begin{center}
\includegraphics[width=0.32\textwidth]{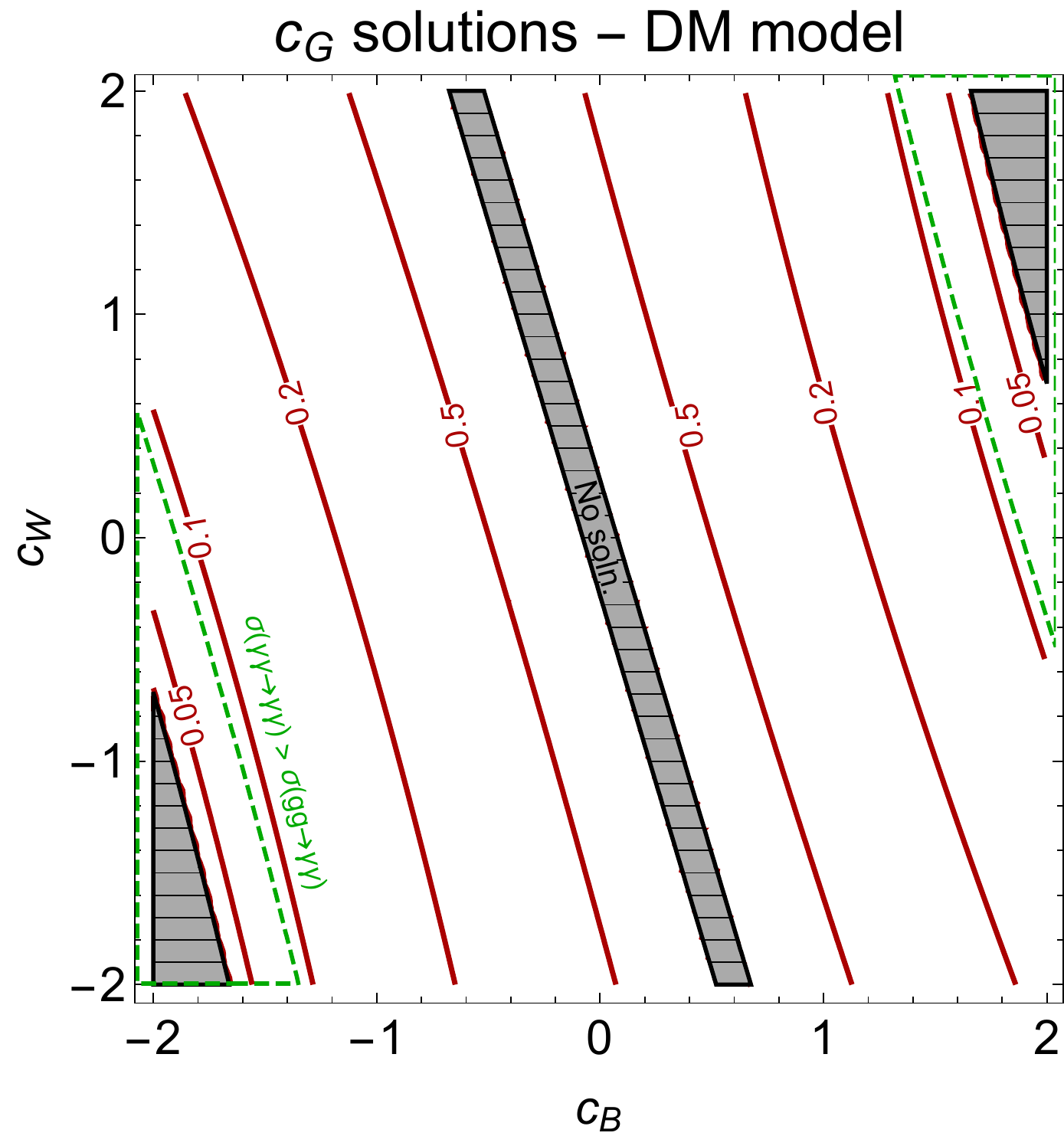}
\includegraphics[width=0.32\textwidth]{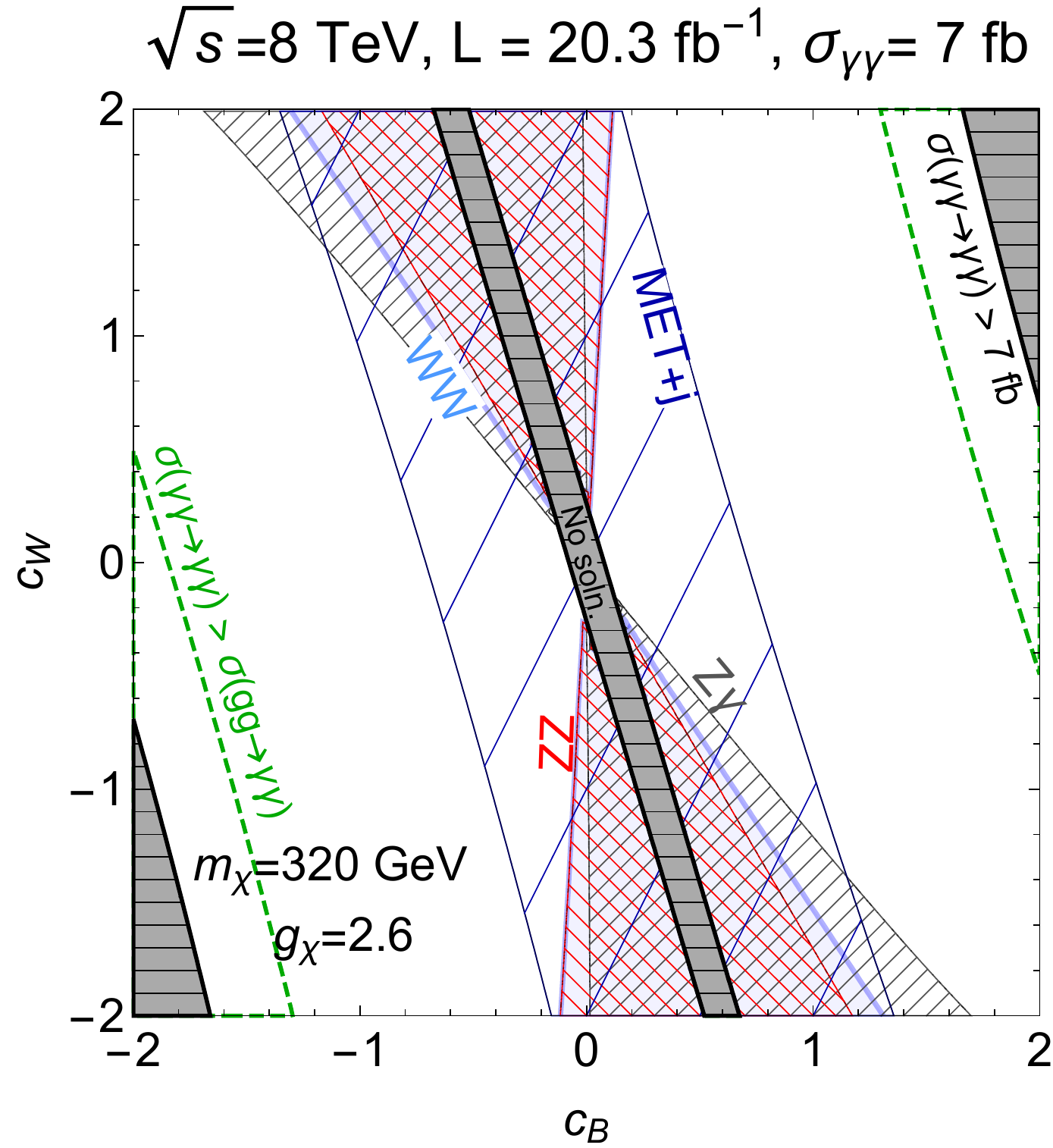}
\includegraphics[width=0.32\textwidth]{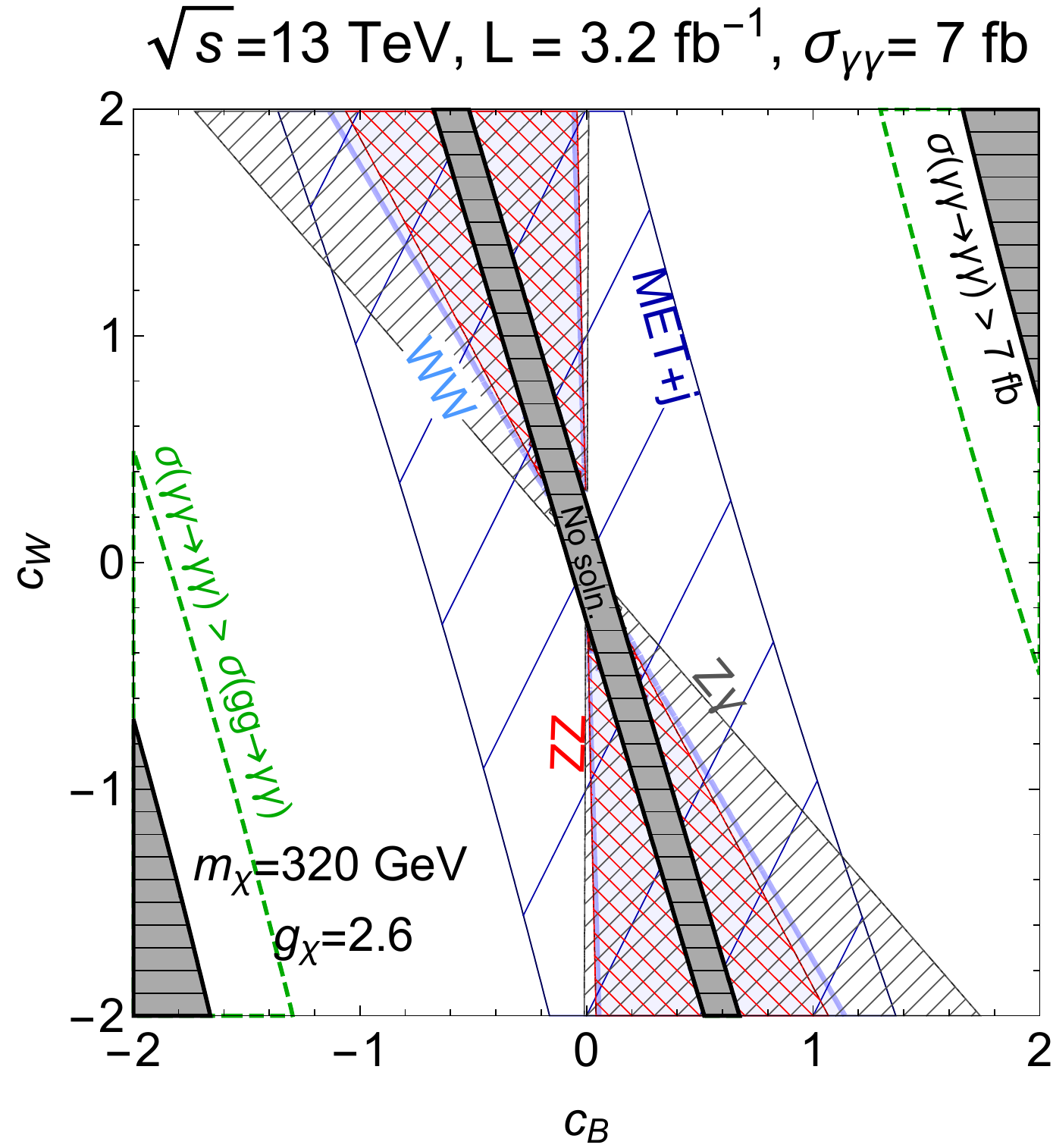}
\includegraphics[width=0.32\textwidth]{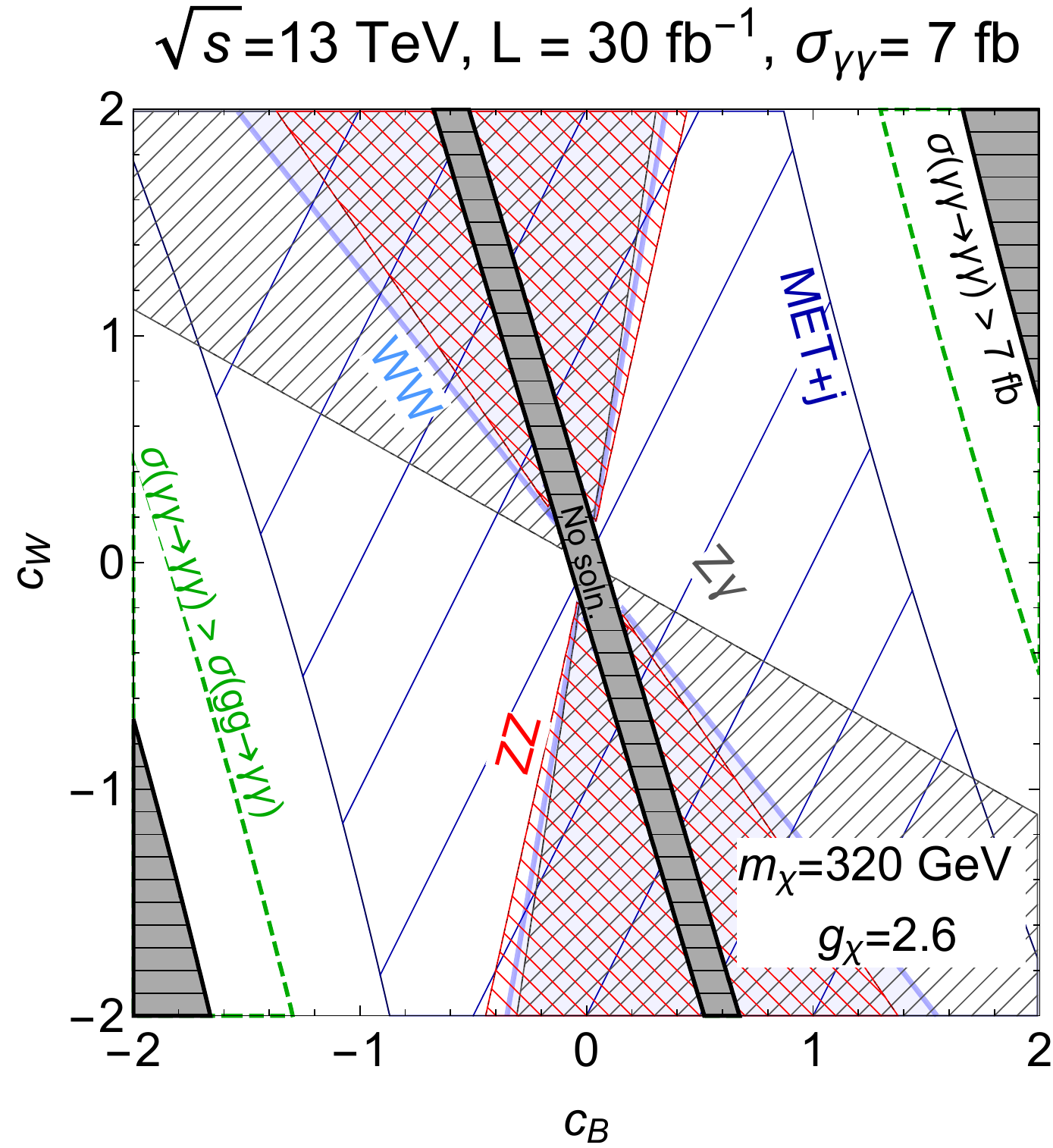}
\includegraphics[width=0.32\textwidth]{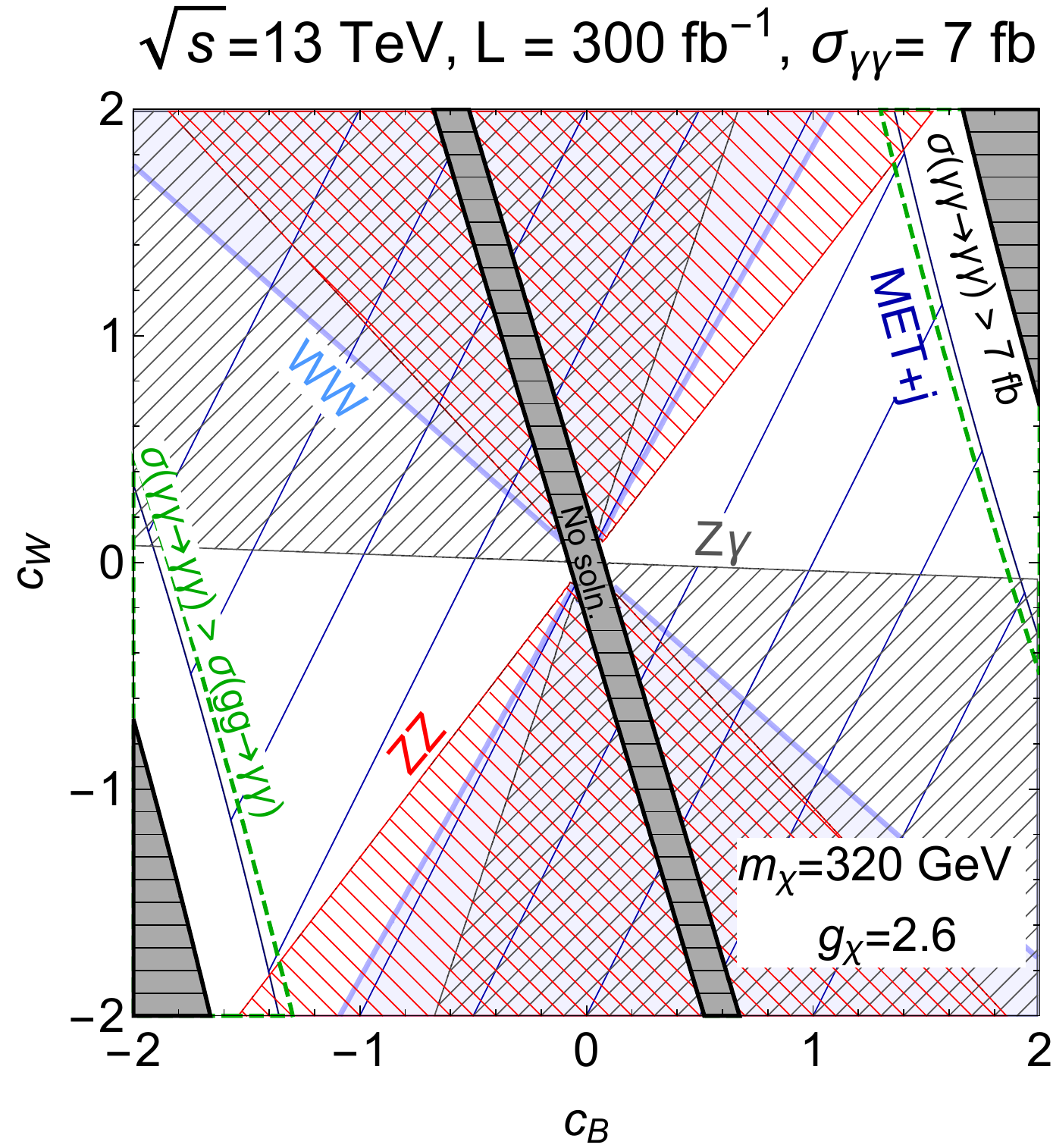}
\caption{Analysis of the ($c_B$, $c_W$) plane in the dark matter model parameter space. 
We apply the bounds from the 8 and 13~TeV runs of the LHC as well as present several projections of bounds at future luminosities.
The dark gray shaded regions are where there is no solution for $c_G$ which can accommodate the diphoton excess. Other shaded regions are excluded by particular searches labeled on the plots.}
\label{fig:dmmodel} 
\end{center}
\end{figure}

Figure~\ref{fig:vanillamodel} does not show the 8~TeV bound on $\gamma \gamma$ final states. In the region where gluon fusion dominates, this bound is automatically satisfied since gluon luminosity increases by a factor of $4.7$, and hence a $\sigma_{\gamma\gamma}(13~\text{TeV})=7~\text{fb}$ corresponds to $\sigma_{\gamma\gamma}(8~\text{TeV})=1.49~\text{fb}$, just below the LHC 8~TeV bound. In the photon-fusion dominated regions the argument is less straightforward. Using the NN23LO1 PDF in MG5 the enhancement factor from $8$ to $13$ TeV in photon fusion is approximately $2$ and hence the photon-fusion dominated regions would not be compatible with LHC 8~TeV constraints. {Given the on-going discussion in the literature about the exact value of the enhancement factor~\cite{Harland-Lang:2016qjy}, it is still possible that photon-fusion is eventually a viable option \cite{Csaki:2015vek,Csaki:2016raa,Harland-Lang:2016qjy,Abel:2016pyc}. Thus, given the large uncertainties in such estimate, conservatively we do not impose any extra bound on such regions from the LHC 8~TeV $\gamma \gamma$ final state searches. 
An ATLAS study of the jet multiplicity distribution in the diphoton events seems to show that the data
favors production processes with a small number of accompanying jets~\cite{ATLAS-CONF-2016-018}, hence consistent with dominant photon-fusion. For all of the above reasons, we choose to simply denote the region with a dashed green line, and remain agnostic on whether it is viable or not.

We proceed to investigate the bounds which are imposed by the LHC 8~TeV searches of resonances 
in the $ZZ$, $Z\gamma$, and $WW$ final states. The results for the 8~TeV limits are displayed in the second top panel of Fig.~\ref{fig:vanillamodel}, where as usual on every point of the plane we have solved for $c_G$ in order
to get $\sigma_{\gamma \gamma}(13\textrm{ TeV})=7$~fb. 
The signal cross section in electroweak boson final states, once the signal yield in $\gamma \gamma$ is imposed, 
is only a function of the ratio $c_B/c_W$, 
which controls the relative size of the branching ratios.\footnote{Note, however, that what we are imposing is a
signal cross section in $\gamma \gamma$ at 13~TeV. 
In the transition from the gluon-fusion to the photon-fusion regime, the corresponding 8~TeV $\gamma\gamma$ 
signal strength changes since the $8\textrm{ TeV}/13\textrm{ TeV}$ ratio of the gluon and photon luminosity is different. 
This effect is not visible in the shape of the regions excluded by the 8~TeV searches since effectively they always lie inside the region
dominated by gluon-fusion.}
As a consequence, the excluded region for each signature has the shape of a symmetric triangular angular slice in the $(c_B,\,c_W)$ plane. 
The strongest constraints come from the $ZZ$ and $Z\gamma$ final states. The $WW$ limit instead provides inferior exclusion power for regions   
already bounded by the other searches. 
The white region is compatible with all existing LHC 8~TeV constraints and fits the 13~TeV diphoton excess.

The remaining panels of Fig.~\ref{fig:vanillamodel} show the LHC 13~TeV reach with increasing luminosity up to $3000\textrm{ fb}^{-1}$.
It is interesting to observe that the $Z \gamma$ limit at $3.2 \textrm{ fb}^{-1}$ is essentially equivalent to the 8~TeV bound, 
while the $WW$ and the $ZZ$ are slightly
weaker. Increasing the luminosity reduces the allowed parameter regions, resulting in a tiny remaining portion at $\mathcal{L}=3000\textrm{ fb}^{-1}$. The result suggests that, if the diphoton excess is confirmed, a complementary signature in weak boson final states is highly likely to be discovered in the coming years. Notice that the projection of the 8~TeV combined  $ZZ$ limit we obtained in Sec.~\ref{sec:setup} plays a crucial role in closing almost entirely the allowed parameter region at the high luminosity LHC.



\subsection{The dark matter model: $g_{X}\neq0 , g_{f}=0$}
\label{sec:dmmodel}


Current ATLAS results favor the interpretation of the diphoton excess in terms of a resonance with a relatively large width 
($i.e.$ $\Gamma/M \sim 5\%$). The large width cannot be explained by decays to gluons and photons alone. Unitarity and the existing di-jet bounds exclude the coupling sizes necessary to generate the large width \cite{Knapen:2015dap}, suggesting that a wide 750 GeV resonance would have to decay to other states as well. As no new charged particles with mass $\sim O(100 \GeV)$ have been observed at the LHC, it is reasonable to consider that the large resonance width can be explained by decays to new invisible particles. Decays of the 750 GeV resonance to neutral states are conceptually very interesting, as non-SM massive particles with no electric charge are natural candidates for dark matter. 

Reference~\cite{Backovic:2015fnp,Barducci:2015gtd, Mambrini:2015wyu,D'Eramo:2016mgv} already considered scenarios in which a scalar $S$ with mass of 750~GeV is allowed to decay to dark matter.
A generic feature appears in most models which explain the large width of $S$ via decays to dark matter: once the values for the decay width and dark matter relic density are fixed, the parameters of the dark sector ($m_X$, $g_X$) are fully determined. For instance, in cases where dark matter is a Dirac fermion coupling to a pure scalar $S$, a large $S$ width and dark matter relic density predict $m_X \approx 300 \GeV$, $g_X \approx 2$.\footnote{The values of the fixed parameter point can change based on the assumptions on the spin and CP properties of dark matter and $S$.}

As an illustration of the LHC prospects to probe the class of the dark matter models for the 750 GeV resonance, here we will consider a benchmark point from Ref.~\cite{Backovic:2015fnp} which is allowed by the current astro-physical and collider constraints:
 $$
m_X = 320 \GeV, \,\,\,\,\,\,\, g_X = 2.6 \,.	\nn
 $$

The first panel of Fig.~\ref{fig:dmmodel}} shows in solid red contours the $c_G$ values necessary to explain the diphoton excess in the dark matter model. The required values of $c_G$ at a fixed $(c_B, c_W)$ are significantly higher compared to the vanilla scenario of the previous section. The reason for a larger $c_G$ stems from the fact that in our dark matter model ${\rm Br}(S\rightarrow X\bar{X}) \approx 1$, requiring larger $c_G$ couplings to compensate for a smaller ${\rm Br}(S\rightarrow gg)$. Notice also that the photon fusion contribution to the $S$ production becomes dominant only for $c_B, c_W \gtrsim 2$. 

The remaining panels of Fig.~\ref{fig:dmmodel} show the results of the current LHC exclusion of the dark matter model parameter space as well as the future prospects. The main difference compared to the vanilla benchmark model of the previous section is that allowing the 750~GeV scalar decays to dark matter introduces constraints from searches in channels with large missing energy, of which we consider MET+$j$. 
The $WW$, $ZZ$, and $\gamma Z$ results constrain the same regions of the parameter space as in the case of the vanilla model, while the MET+$j$ channel typically provides the strongest limits, except in the corners of large $(-c_B, c_W)$. We find that current 8~TeV and 13~TeV results exclude $c_B$ values in the range of $ |c_B| \lesssim 0.2$ for $c_W = 2$, up to values of $|c_B| \lesssim 1.7$ for $c_W = -2$. Future LHC results at 13~TeV will be able to exclude a majority of the parameter space with as little as $30\,\fb^{-1}$ of data, while with 
$300\,\fb^{-1}$ only the regions of parameter space in which photon-fusion dominates will not be ruled out by MET+$j$.



\subsection{The ``top-philic" model: $g_{X}=0 , g_{f}\neq0$}

As a final concrete example of the diphoton models, we discuss the case in which the new scalar resonance also
couples to SM fermions, $i.e.$ we set $g_{X}=0$, with non-vanishing $g_f$ in Eq.~\eqref{Lag_full}.\footnote{Note that the coupling of $S$ to the SM fermions will generate extra contributions to the effective operator between $S$ and the gauge bosons
(see for instance \cite{Bellazzini:2015nxw} for the case of a pseudoscalar coupled to gauge bosons and top quark).
However, since we consider the same suppression scale $\Lambda$ for all dimension five operators, and all couplings $(c_B,c_W,c_G)$ and $g_f$ of order $O(1)$, 
such loop induced contributions will be typically subleading on the parameter space under study.}

Among the various couplings to the SM fermions, 
the dominant coupling is to the top quark, justifying the title ``top-philic''. In particular, the coupling of $S$ with the
top quark will  induce a sizable decay width of $S$ into $\bar t t$ pairs (see Appendix \ref{sec:ana_formulae}), 
which now constitutes the dominant decay mode of the scalar resonance. As a consequence, in order to
obtain the desired signal strength in the $\gamma \gamma$ final state at 13~TeV, the production cross section for $S$ should be quite sizeable compared to the vanilla model of Sec.~\ref{sec:vanilla}.
The top-philic model is then similar to the dark matter model studied before, 
where the dominant invisible decay has been now substituted by a dominant
decay into top-antitop pairs. 

\begin{figure}[!]
\begin{center}
\includegraphics[width=0.32\textwidth]{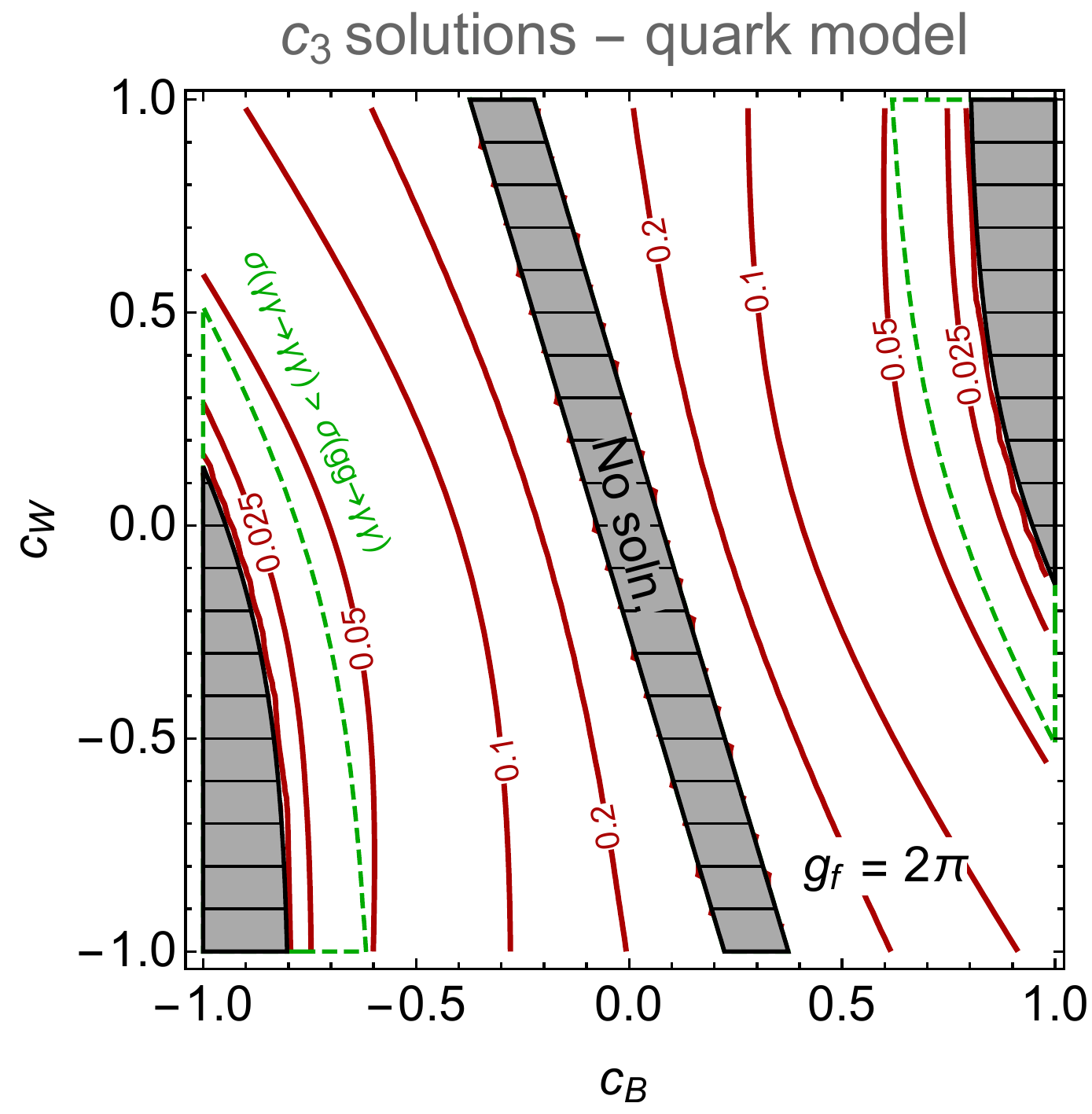}
\includegraphics[width=0.32\textwidth]{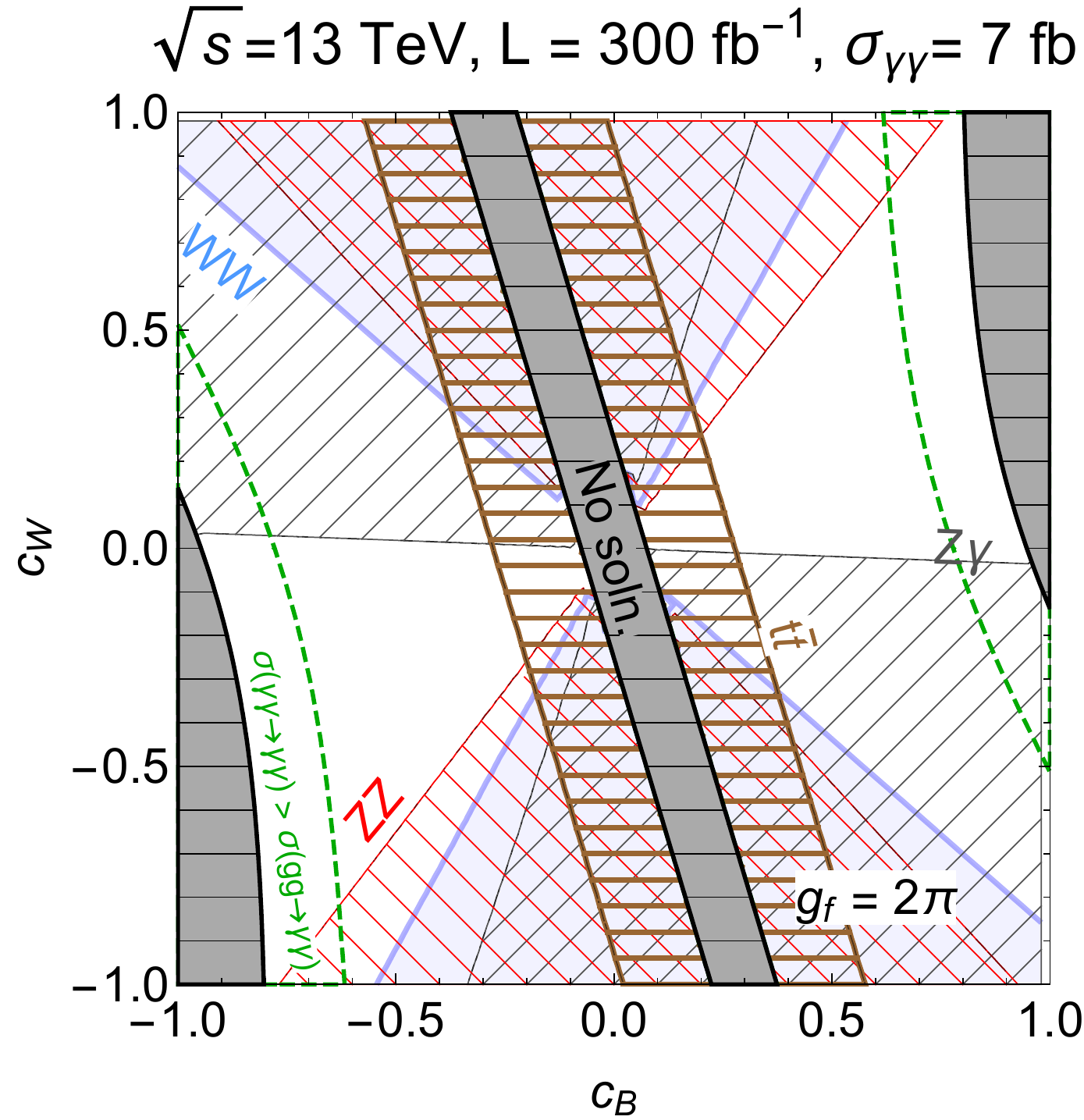}
\includegraphics[width=0.32\textwidth]{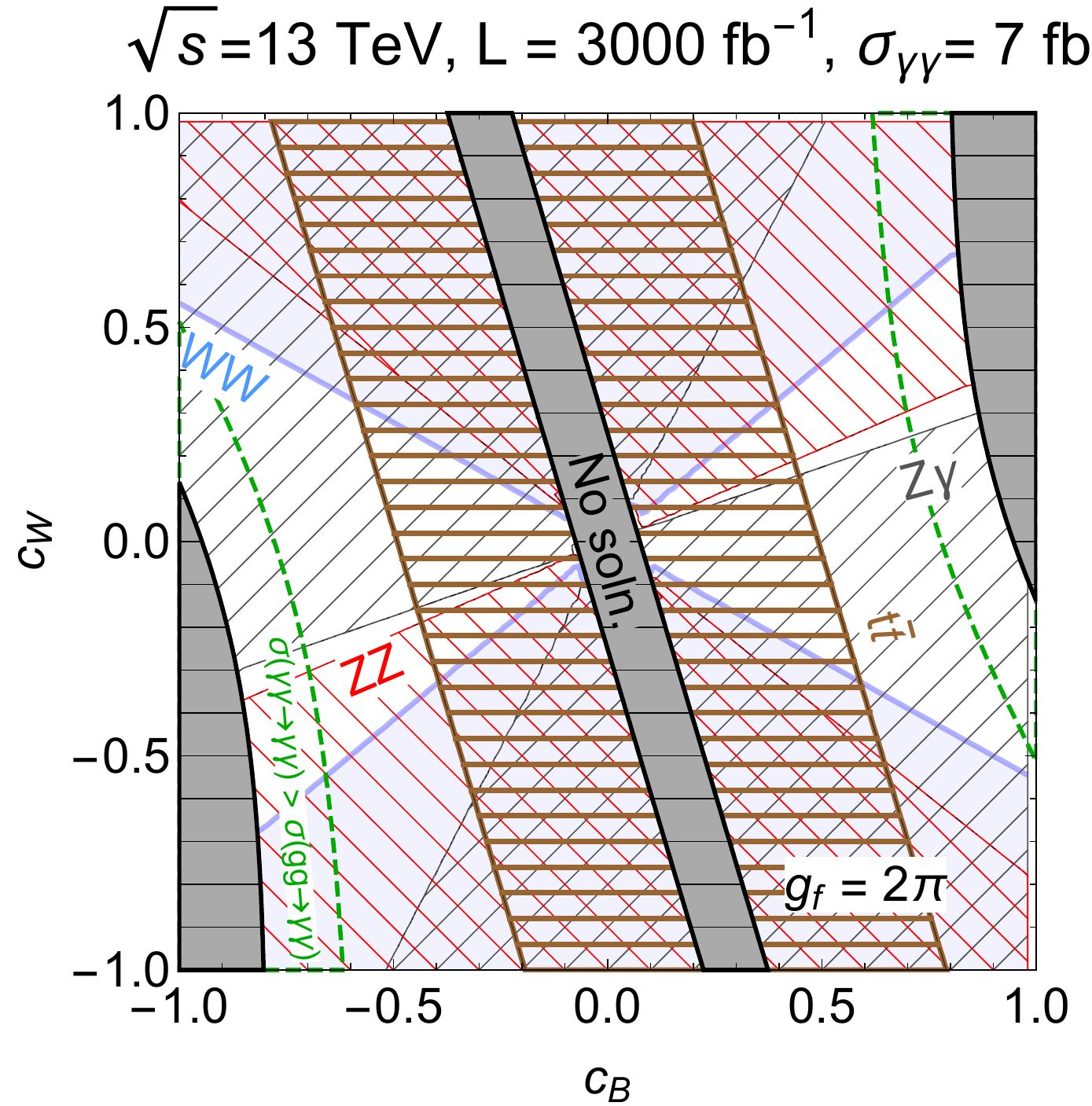}
\caption{Analysis of the the top-philic model parameter space. The dark gray shaded regions are where there is no $c_G$ solution for 
the diphoton excess. Other shaded regions are excluded by particular searches labeled on the plots. 
The left and right panels correspond to two benchmark values of $g_f$.}
\label{fig:SMfermmodel} 
\end{center}
\end{figure}

We show the results of our analysis in the case of the top-philic model in Fig.~\ref{fig:SMfermmodel} for one representative value $g_f = 2\pi$ at the high luminosity LHC. Indeed, note that since the coupling to SM fermions are
suppressed by a factor $m_q/\Lambda$ only large values of $g_f$ will induce interesting effects.
The only final state which distinguishes the top-philic model from the vanilla scenario of Sec.~\ref{sec:vanilla} is $t\bar{t}$. The brown shaded regions in Fig.~\ref{fig:SMfermmodel} illustrate the regions of the parameter space the future $t \bar t$ resonance searches will be able to probe. The example we show in Fig.~\ref{fig:SMfermmodel} suggests that the top-philic model can be probed with high luminosity LHC only in the regime of $g_f \gtrsim \pi$.
In the large $g_f$ scenario, the addition of the $t \bar t$ channel to the usual electroweak boson searches essentially covers the entire parameter space that we considered with 3000 $\fb^{-1}$ of integrated luminosity at 13 TeV LHC.
Note that the presence of a large coupling to the top quark, 
pushes the photon-fusion dominated region further to larger values of $c_B$ compared to the vanilla model 
of Sec.~\ref{sec:vanilla}. The reason is that a large coupling to SM fermions implies a small $\textrm{Br}(S\rightarrow\gamma\gamma)$, 
resulting in the need of larger production cross section to accommodate the excess, that can essentially be obtained only 
via gluon fusion in the range of $c_B$ under consideration.

\section{Summary}
\label{sec:sum}
In this paper we have explored the LHC 13~TeV reach for models capable of explaining the diphoton excess at 750 GeV.
As illustrative example we have considered a simple model with a scalar resonance coupled to SM gauge bosons, 
a dark matter candidate, and the SM quarks. 
We took into account gluon-fusion as well as photon-fusion as production mechanisms at the LHC.
The requirement of generating the correct cross section in $\gamma \gamma$ final state at 13~TeV 
 imposes relations among the model parameters.  
We have studied the correlated signatures that can arise in such scenarios, including  final states with di-bosons, jet plus missing energy, 
and $t \bar t$ resonance searches, in order to further constrain the parameter space of the model 
and establish the exclusion reach of the LHC 13~TeV. 

Our findings indicate that correlated LHC searches can exclude most of the relevant parameter space of a broad class of diphoton models during the second run of the 
LHC.
The ``vanilla" model (where the scalar resonance is coupled only to SM gauge bosons with dimension five operators) can
be almost completely covered by associated signals in di-bosons with $3000\, \fb^{-1}$ of integrated luminosity.
Concerning models where $S$ is a portal to a dark sector, we show that the mono-jet searches are able to corner the model with as little as $30 \textrm{ fb}^{-1}$.
Finally, for models where the scalar resonance couples to SM quarks, the signature in the $t \bar t$ final state could provide
a handle on distinguishing such scenarios from the ``vanilla" model. However, in order for the signal in $t\bar{t}$ to be accessible,  sizeable couplings of quarks to $S$ are required, as well as integrated luminosity of at least $300 \textrm{ fb}^{-1}$.

It would be interesting to extend our work to more exotic scenarios that can explain the diphoton excess,
including $e.g.$ non-resonant production, collimated photons, and models with non trivial coupling with the Higgs boson.
The procedure we have adopted in this paper to compare with extrapolated LHC 13 TeV limits could be extended also to such scenarios. If the di-photon excess is confirmed, it becomes of utmost importance to explore the full set of correlated signatures expected to appear in the ongoing run of the LHC.

\paragraph{Note added:}
During the final stages of this work, Refs.~\cite{Sato:2016hls} and \cite{No:2016htu} appeared.  
Both references studied the LHC prospects for exclusion of a simplified diphoton resonance model analogous to the scenario 
we study in Section~\ref{sec:vanilla}, and obtained results which are in agreement with ours. 
Compared to Refs.~\cite{Sato:2016hls} and \cite{No:2016htu},
our analysis in 
Section~\ref{sec:vanilla} also discusses the production mechanism 
for the resonance, including gluon and photon fusion. 

\bigskip

\noindent \textbf{Acknowledgments}
\medskip

\noindent We would like to thank A. Goudelis, K. Kowalska, D. Redigolo, and F. Sala for useful discussions. 
M.B. is supported by a MOVE-IN Louvain Cofund grant. S.K. is supported by the ``New Frontiers" program of the Austrian Academy of Sciences. A.M. is supported by the Strategic Research Program High Energy Physics and the Research Council of the Vrije Universiteit Brussel. M.B. and A.M. are also supported in part by the Belgian Federal Science Policy Office through the Interuniversity Attraction
Pole P7/37. M.S. is supported in part by the European Commission through the ``HiggsTools'' Initial Training Network PITN-GA-2012-316704.  
%



\appendix

\section{Limit extrapolation}
\label{sec:limit}

In order to project the LHC 8~TeV limits to 13~TeV,  we employ a simple extrapolation algorithm, similar to Refs.~\cite{Thamm:2015zwa,Buttazzo:2015bka}. 
We begin with the assumption that the 8~TeV and 13~TeV resonance searches are characterized by acceptances and event selection efficiencies which are roughly equal. 

The $CL_s$ test statistics employed by the experimental collaborations to determine the 95\%~C.L. 
upper bounds on the cross section times branching ratio is a constant at different luminosities and center of mass energies:
\begin{equation}
CL_s\left(s,\mathcal{L},M\right)\equiv CL_s\left(S^{\textrm{max}}(s,\mathcal{L},M),\,B(s,\mathcal{L},M) \right)=95\%=\textrm{const}\,,\label{cls}
\end{equation}
where $S^{\textrm{max}}$ is the upper bound on the number of signal events and $B$ is the expected or observed background,
$s$ is the center of mass energy, $\mathcal{L}$ the integrated luminosity, and $M$ the invariant mass bin. 

Assuming that the background is dominated by a single initial state production mode 
(which is a decent approximation in most cases) we can write at any given $\mathcal{L}$ and $M$:
\be
	B (s, \mathcal{L}, M)  = r_{ij}(M, s) \times B (s^0, \mathcal{L}, M), \label{lim2}
\ee
where $r_{ij}$ is the parton luminosity ratio and $i, j$ stand for quarks and gluons.

Inserting this in Eq.~(\ref{cls})
\begin{eqnarray}
CL_s\left(S^{\textrm{max}}(s,\mathcal{L},M),\,B(s,\mathcal{L},M) \right)&=&CL_s\left(S^{\textrm{max}}(s,\mathcal{L},M),\,r_{ij}(M, s) \times B (s^0, \mathcal{L}, M)\right)\nonumber\\
 &=&CL_s\left(S^{\textrm{max}}(s,\mathcal{L},M),\, r_{ij}(M, s) \times \frac{\mathcal{L}}{\mathcal{L}^0}\,B (s^0, \mathcal{L}^0, M)\right)\nonumber\\
 &=&CL_s\left(S^{\textrm{max}}(s^0,\mathcal{L}^0,M),\,B(s^0,\mathcal{L}^0,M) \right)\,,\label{array}
\end{eqnarray}
where in the last line we have used the fact the the $CL_s$ is a constant, see Eq.~(\ref{cls}).

In the limit of a large number of events the event ditribution becomes well approximated by a Gaussian, so that the equality 
between the second-to-last and last line of Eq.~(\ref{array}) can be written as
\begin{equation}
\frac{S^{\textrm{max}}(s,\mathcal{L},M)}{ \left[r_{ij}(M, s) \times \frac{\mathcal{L}}{\mathcal{L}^0}\,B (s^0, \mathcal{L}^0, M)\right]^{1/2}}=
\frac{S^{\textrm{max}}(s^0,\mathcal{L}^0,M)}{B(s^0, \mathcal{L}^0, M)^{1/2}}\,.\label{equo}
\end{equation}

Moreover, because of our initial assumption that the efficiencies and acceptances are the same, 
$S^{\textrm{max}}(s,\mathcal{L},M)$ scales as $\mathcal{L}\times\sigma^{\rm max} (s, M)$, so that 
one solves Eq.~(\ref{equo}) to get 
\be
	\sigma^{\rm max} (s, M) \approx  \sqrt{r_{ij}(M,s)} \times \sqrt{\frac{\mathcal{L}^0}{\mathcal{L}}}  \times \sigma^{\rm max} (s^0, M)\,.  \label{eq:extrapol}
\ee

Equation~\eqref{eq:extrapol} represents our ``master formula'' for limit extrapolations.
The parton luminosity ratios $r_{ij} (M, s) $ have been previously calculated in Ref.~\cite{jsterling}. For completeness, here we give a numerical polynomial fit to the parton luminosity ratios for $gg$, $\sum (qq + q\bar{q})$ and $qg$ initial states, valid in the range of $M = 50 - 4000 \GeV$:
\begin{eqnarray}
	r_{gg}(x) &\approx& 1.6 + 6.3\times10^{-3} \,x - 7.9\times10^{-6}  \,x^2 + 8.8\times10^{-9} \,x^3 \nonumber \\
			& &- 3.7\times 10^{-12} \,x^4 + 6.6\times10^{-16}\, x^5\, , \nonumber \\
	r_{qq}(x) & \approx& 1.7 - 2.5 \,x + 6.0\times 10^{-6}\, x^2 - 8.8\times10^{-9}\, x^3 \nonumber \\
			& &+  6.2\times10^{-12}\, x^4 - 1.9\times10^{-15} \,x^5 + 2.3\times10^{-19} \,x^6\,, \nonumber \\
	r_{qg}(x) &\approx&  1.3 + 7.1\times 10^{-3} \,x - 1.2\times 10^{-5} \,x^2 \nonumber \\
			& &+ 1.1\times 10^{-8} \,x^3 - 4.1 \times 10^{-12} \,x^4 + 5.7\times10^{-16} \,x^5\,,
\end{eqnarray}
where $x\equiv M/\GeV$.

\begin{figure}[!]
\begin{center}
\includegraphics[width=0.4\textwidth]{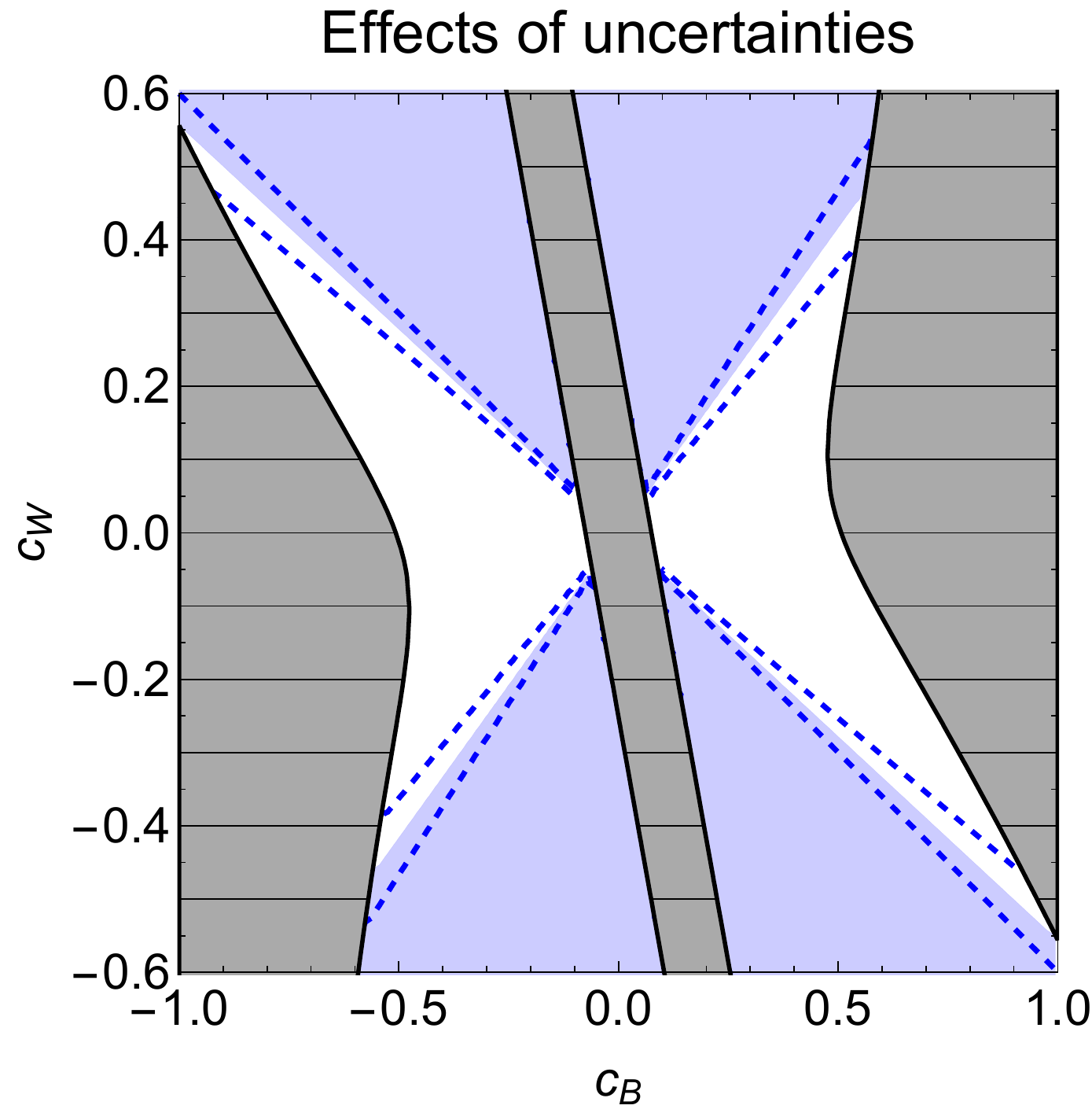}
\caption{Effects of uncertainties of the limit extrapolation procedure on the exclusion regions of the model parameter space. The blue, shaded region shows the portion of the ``base model'' parameter space excluded by the extrapolated $WW$ limit at 13 TeV with $\mathcal{L} = 3000 \,\fb^{-1}$, with the dashed lines showing the position of the excluded region edges if the limit on the cross section was $\pm 20\%$ different. The gray shaded regions represent the parameter space where either no-viable solution for $c_G$ can be found to accommodate the excess or the predicted diphoton cross section at 13 TeV is too big.}
\label{fig:uncertainty} 
\end{center}
\end{figure}

We find that when used to extrapolate the expected 8~TeV limits, the extrapolation formula of Eq.~\eqref{eq:extrapol} gives results which are within $\sim 20\%$ from the true expected limits at 13 TeV. 
In order to validate the procedure, we have compared the results using Eq.~\eqref{eq:extrapol} to a number of already public ATLAS results from 13~TeV. Table~\ref{tab:maremma} shows the results. The largest error in our limit extrapolation is $\sim 28-29\%$, in the case of the $\gamma\gamma$ and $Z\gamma$ searches. 
This is mostly due to the fact that ATLAS does not provide for those searches the efficiencies for all bins, 
and a knowledge of the latter is required to extrapolate the cross section bound for the fiducial cross section bound.
The average error is about 10\%. The uncertainty in the limit extrapolation does not strongly affect our results on the parameter space exclusion. Figure~\ref{fig:uncertainty} illustrates the result in case of the $WW$ cross section, extrapolated from the $8 \TeV$ limit to 13\,TeV with $\mathcal{L} = 3000\,\fb^{-1}$. The blue, shaded region shows the excluded parameter space, while the dashed regions show where the edge of the exclusion would lie if the maximal cross section was $\pm 20 \%$ different. 

Although Eq.~\eqref{eq:extrapol} gives reasonably accurate results in many cases, it is important to point out where it fails. 
If the event reconstruction and selection efficiencies and acceptances differ significantly between 8 TeV and 13 TeV,  
Eq.~\eqref{eq:extrapol} can result in errors larger than $20 \%$. 
The approximation is also not accurate when the 8~TeV expected background 
is a number of the order of a few units, so that the event distribution is not well approximated by a Gaussian, 
but rather presents a longer tail.

Another scenario in which the extrapolation of Eq.~\eqref{eq:extrapol} fails are non-resonance searches ($e.g.$ MET+$j$) 
or searches for broad resonances. In cases where the signal cross section is not distributed mostly in a narrow range of invariant masses
(such as in the case of a narrow resonance), it is inappropriate to use a parton luminosity ratio evaluated at a single $M$. 
Instead, an integral value over the parton luminosities is more appropriate, as the signal cross section will be distributed over a wider 
range of $M$. 

\begin{table}
\footnotesize
\begin{center}
\begin{tabular}{cccccccc}
F.S. & Ref. & $M_{\rm res} $[GeV] & I.S. &$\sigma_{\rm exp}^{\rm max},\,\mathcal{L} (8 \TeV)$ & $\sigma_{\rm exp}^{\rm max},\,\mathcal{L}(13 \TeV)$ & $\sigma_{\rm ext}^{\rm max}$(13 TeV)& \% diff.\\
\hline
\multirow{2}{*}{$Zh$}&\multirow{2}{*}{8 TeV \cite{Aad:2015wra}}& 300 &\multirow{2}{*}{$gg$} & 220 fb, 20.3 $\fb^{-1}$ & 1250 fb, 3.2 $\fb^{-1}$ & 952 fb& -27 \% \\
&& 400 & & 92 fb, 20.3 $\fb^{-1}$ & 500 fb, 3.2 $\fb^{-1}$& 423 fb & -17 \% \\
 &\multirow{2}{*}{13 TeV \cite{ATLAS-CONF-2016-015}}& 750 & & 16 fb, 20.3 $\fb^{-1}$ & 73 fb, 3.2 $\fb^{-1}$& 87 fb & +17 \% \\
&& 1000 & & 10 fb, 20.3 $\fb^{-1}$ & 50 fb, 3.2 $\fb^{-1}$& 61 fb & +20 \% \\
\hline
\multirow{2}{*}{$Z\gamma$}&\multirow{2}{*}{8 TeV \cite{Aad:2014fha}}& 400 &\multirow{2}{*}{$qq$} & 0.5 fb, 20.3 $\fb^{-1}$ & 2.3 fb, 3.2 $\fb^{-1}$ &1.8 fb& -24 \% \\
 & & 750 & & 0.2 fb, 20.3 $\fb^{-1}$ & 1.2 fb, 3.2 $\fb^{-1}$& 0.9 fb & -29 \% \\
 &13 TeV \cite{ATLAS-CONF-2016-010} & 1600 & & 0.1 fb, 20.3 $\fb^{-1}$ & 0.6 fb, 3.2 $\fb^{-1}$& 0.6 fb & 0 \% \\
\hline
\multirow{2}{*}{$ll$}&\multirow{2}{*}{8 TeV \cite{Aad:2014cka}}& 500 &\multirow{2}{*}{$qq$} & 3.2 fb, 20.4 $\fb^{-1}$ & 11 fb, 3.2 $\fb^{-1}$ &12 fb & +9 \% \\
 & & 750 & & 1.2 fb, 20.4 $\fb^{-1}$ & 4.8 fb, 3.2 $\fb^{-1}$ & 4.9 fb & +2 \% \\
 & 13 TeV \cite{ATLAS-CONF-2015-070} & 1500 & & 0.4 fb, 20.4 $\fb^{-1}$ & 1.6 fb, 3.2 $\fb^{-1}$ & 1.9 fb & +17 \% \\
\hline
\multirow{2}{*}{$ZZ$}&\multirow{2}{*}{8 TeV~\cite{Aad:2014xka}}& 750 &\multirow{2}{*}{$qq$} & 48 fb, 20.3 $\fb^{-1}$ & 200 fb, 3.2 $\fb^{-1}$ & 197 fb& -2 \% \\
&& 1000 & & 19 fb, 20.3 $\fb^{-1}$ & 105 fb, 3.2 $\fb^{-1}$& 85 fb & -21 \% \\
$(llqq) $& 13 TeV \cite{ATLAS-CONF-2015-071} & 2000 & & 6.0 fb, 20.3 $\fb^{-1}$ & 38 fb, 3.2 $\fb^{-1}$& 41 fb & +8 \% \\
\hline
\multirow{2}{*}{$hh$}&\multirow{2}{*}{8 TeV \cite{Aad:2015uka}}& 600 &\multirow{2}{*}{$gg$} & 22 fb, 19.5 $\fb^{-1}$ & 110 fb, 3.2 $\fb^{-1}$ &110 fb& 0 \% \\
& & 800 & & 9 fb, 19.5 $\fb^{-1}$ & 60 fb, 3.2 $\fb^{-1}$& 49 fb & -20 \% \\
& 13 TeV~\cite{ATLAS-CONF-2016-017} &1400 & & 3.9 fb, 19.5 $\fb^{-1}$ & 22 fb, 3.2 $\fb^{-1}$& 28 fb & +24 \% \\
\hline
\multirow{2}{*}{$\gamma\gamma$}&\multirow{2}{*}{8 TeV \cite{Aad:2015mna}}& 500 &\multirow{2}{*}{$gg$} & 4.1 fb, 20.3 $\fb^{-1}$ & 22 fb, 3.2 $\fb^{-1}$ & 20 fb& -10 \% \\
 & &750 & & 2.0 fb, 20.3 $\fb^{-1}$ & 8.2 fb, 3.2 $\fb^{-1}$& 10.9 fb & +28 \% \\
 & 13 TeV \cite{ATLAS-CONF-2015-081} & 1500 & & 0.5 fb, 20.3 $\fb^{-1}$ & 3.9 fb, 3.2 $\fb^{-1}$& 3.8 fb & -3 \% \\ 
\hline
\end{tabular}
\caption{Validations of the limit extrapolation procedure from 8 to 13~TeV. In the table F.S. stands for decay ``final state" and I.S. for
production ``initial state". 
We extracted the expected limits from the corresponding references listed in the table. Percent difference is defined as 
$2 [\sigma^{\rm max}_{\rm ext}(13 \TeV)  - \sigma^{\rm max}_{\rm exp}(13 \TeV)] / [\sigma^{\rm max}_{\rm ext} (13 \TeV) +
 \sigma^{\rm{max}}_{\rm exp}(13 \TeV)]$. Extrapolations are accurate within a $\sim 20\%$ margin. 
The only shown exception involves the 750~GeV bin of the $\gamma\gamma$ search, as~\cite{ATLAS-CONF-2015-081} 
does not provide a detailed account of the acceptances/efficiencies in all bins, which are necessary
when comparing the ``fiducial' cross section to the physical cross section.}
\label{tab:maremma}
\end{center}
\end{table} 



\section{Analytical form of the decay widths}
\label{sec:anfor}

In this appendix we report the analytic formulas for the partial decay widths of the resonance. 
The following expressions were used for the analysis
discussed in the main body of the paper.
\bee
&&
\Gamma[S \to \gamma \gamma]=\frac{(c_B \cos^2 \theta_W +c_W \sin^2 \theta_W)^2 m_{S}^3}{4 \pi \Lambda^2} \\
&&
\Gamma[S \to g g]=\frac{2 c_G^2 m_{S}^3}{\Lambda^2 \pi} \\
&&
\Gamma[S \to ZZ]=\frac{(c_B \sin^2 \theta_W +c_W \cos^2 \theta_W)^2 m_{S}^3}{4 \pi \Lambda^2} \left(1-4 \frac{m_Z^2}{m_{S}^2}+6\frac{m_Z^4}{m_{S}^4}\right) \sqrt{1-\frac{4 m_Z^2}{m_{S}^2}} \\
&&
\Gamma[S \to Z\gamma]=\frac{(\sin \theta_W \cos \theta_W(c_B-c_W))^2 m_{S}^3}{2 \pi \Lambda^2} \left(1-\frac{m_{Z}^2}{m_{S}}\right)^3\\
&&
\Gamma[S \to W^+ W^-]=\frac{c_W^2 m_{S}^3}{2 \pi \Lambda^2} \left(1-4 \frac{m_W^2}{m_{S}^2}+6\frac{m_W^4}{m_{S}^4}\right) \sqrt{1-\frac{4 m_W^2}{m_{S}^2}} \\
&&
\Gamma[S \to \bar XX] = \frac{g_{X}^2 m_{S}}{8 \pi }\left(1-\frac{4 m_{X}^2}{m_{S}^2} \right)^{3/2} \\
&&
\Gamma[S \to \bar q q] = \frac{3 g_{f}^2 m_{S} m_q^2}{8 \pi \Lambda^2 }\left(1-\frac{4 m_{q}^2}{m_{S}^2} \right)^{3/2}
\eee

\label{sec:ana_formulae}

  
\bibliographystyle{JHEP}

\bibliography{myref}  

\providecommand{\href}[2]{#2}\begingroup\raggedright\begin{thebibliography}{10}

\bibitem{ATLAS-CONF-2015-081}
{\it {Search for resonances decaying to photon pairs in 3.2 fb$^{-1}$ of $pp$
  collisions at $\sqrt{s}$ = 13 TeV with the ATLAS detector}},  Tech. Rep.
  ATLAS-CONF-2015-081, CERN, Geneva, Dec, 2015.

\bibitem{ATLAS-CONF-2016-018}
{\it {Search for resonances in diphoton events with the ATLAS detector at
  $\sqrt{s}$ = 13 TeV}},  Tech. Rep. ATLAS-CONF-2016-018, CERN, Geneva, Mar,
  2016.

\bibitem{CMS-PAS-EXO-15-004}
{\bf CMS Collaboration} Collaboration, {\it {Search for new physics in high
  mass diphoton events in proton-proton collisions at $\sqrt{s} = 13$ TeV}},
  Tech. Rep. CMS-PAS-EXO-15-004, CERN, Geneva, 2015.

\bibitem{CMS-PAS-EXO-16-018}
{\bf CMS Collaboration} Collaboration, {\it {Search for new physics in high
  mass diphoton events in $3.3~\mathrm{fb}^{-1}$ of proton-proton collisions at
  $\sqrt{s}=13~\mathrm{TeV}$ and combined interpretation of searches at
  $8~\mathrm{TeV}$ and $13~\mathrm{TeV}$}},  Tech. Rep. CMS-PAS-EXO-16-018,
  CERN, Geneva, 2016.

\bibitem{Low:2015qho}
I.~Low and J.~Lykken, {\it {Implications of Gauge Invariance on a Heavy
  Diphoton Resonance}},  \href{http://arxiv.org/abs/1512.09089}{{\tt
  arXiv:1512.09089}}.

\bibitem{Kamenik:2016tuv}
J.~F. Kamenik, B.~R. Safdi, Y.~Soreq, and J.~Zupan, {\it {Comments on the
  diphoton excess: critical reappraisal of effective field theory
  interpretations}},  \href{http://arxiv.org/abs/1603.06566}{{\tt
  arXiv:1603.06566}}.

\bibitem{Backovic:2015fnp}
M.~Backovic, A.~Mariotti, and D.~Redigolo, {\it {Di-photon excess illuminates
  Dark Matter}},  {\em JHEP} {\bf 03} (2016) 157,
  [\href{http://arxiv.org/abs/1512.04917}{{\tt arXiv:1512.04917}}].

\bibitem{Barducci:2015gtd}
D.~Barducci, A.~Goudelis, S.~Kulkarni, and D.~Sengupta, {\it {One jet to rule
  them all: monojet constraints and invisible decays of a 750 GeV diphoton
  resonance}},  \href{http://arxiv.org/abs/1512.06842}{{\tt arXiv:1512.06842}}.

\bibitem{Mambrini:2015wyu}
Y.~Mambrini, G.~Arcadi, and A.~Djouadi, {\it {The LHC diphoton resonance and
  dark matter}},  {\em Phys. Lett.} {\bf B755} (2016) 426--432,
  [\href{http://arxiv.org/abs/1512.04913}{{\tt arXiv:1512.04913}}].

\bibitem{D'Eramo:2016mgv}
F.~D'Eramo, J.~de~Vries, and P.~Panci, {\it {A 750 GeV Portal: LHC
  Phenomenology and Dark Matter Candidates}},  {\em JHEP} {\bf 05} (2016) 089,
  [\href{http://arxiv.org/abs/1601.01571}{{\tt arXiv:1601.01571}}].

\bibitem{Franceschini:2015kwy}
R.~Franceschini, G.~F. Giudice, J.~F. Kamenik, M.~McCullough, A.~Pomarol,
  R.~Rattazzi, M.~Redi, F.~Riva, A.~Strumia, and R.~Torre, {\it {What is the
  $\gamma \gamma$ resonance at 750 GeV?}},  {\em JHEP} {\bf 03} (2016) 144,
  [\href{http://arxiv.org/abs/1512.04933}{{\tt arXiv:1512.04933}}].

\bibitem{Falkowski:2015swt}
A.~Falkowski, O.~Slone, and T.~Volansky, {\it {Phenomenology of a 750 GeV
  Singlet}},  {\em JHEP} {\bf 02} (2016) 152,
  [\href{http://arxiv.org/abs/1512.05777}{{\tt arXiv:1512.05777}}].

\bibitem{Gupta:2015zzs}
R.~S. Gupta, S.~Jäger, Y.~Kats, G.~Perez, and E.~Stamou, {\it {Interpreting a
  750 GeV Diphoton Resonance}},  \href{http://arxiv.org/abs/1512.05332}{{\tt
  arXiv:1512.05332}}.

\bibitem{Csaki:2015vek}
C.~Csaki, J.~Hubisz, and J.~Terning, {\it {Minimal model of a diphoton
  resonance: Production without gluon couplings}},  {\em Phys. Rev.} {\bf D93}
  (2016), no.~3 035002, [\href{http://arxiv.org/abs/1512.05776}{{\tt
  arXiv:1512.05776}}].

\bibitem{Csaki:2016raa}
C.~Csaki, J.~Hubisz, S.~Lombardo, and J.~Terning, {\it {Gluon vs. Photon
  Production of a 750 GeV Diphoton Resonance}},
  \href{http://arxiv.org/abs/1601.00638}{{\tt arXiv:1601.00638}}.

\bibitem{Harland-Lang:2016qjy}
L.~A. Harland-Lang, V.~A. Khoze, and M.~G. Ryskin, {\it {The production of a
  diphoton resonance via photon-photon fusion}},  {\em JHEP} {\bf 03} (2016)
  182, [\href{http://arxiv.org/abs/1601.07187}{{\tt arXiv:1601.07187}}].

\bibitem{Abel:2016pyc}
S.~Abel and V.~V. Khoze, {\it {Photo-production of a 750 GeV di-photon
  resonance mediated by Kaluza-Klein leptons in the loop}},  {\em JHEP} {\bf
  05} (2016) 063, [\href{http://arxiv.org/abs/1601.07167}{{\tt
  arXiv:1601.07167}}].

\bibitem{Fichet:2015vvy}
S.~Fichet, G.~von Gersdorff, and C.~Royon, {\it {Scattering light by light at
  750 GeV at the LHC}},  {\em Phys. Rev.} {\bf D93} (2016), no.~7 075031,
  [\href{http://arxiv.org/abs/1512.05751}{{\tt arXiv:1512.05751}}].

\bibitem{Jaeckel:2012yz}
J.~Jaeckel, M.~Jankowiak, and M.~Spannowsky, {\it {LHC probes the hidden
  sector}},  {\em Phys. Dark Univ.} {\bf 2} (2013) 111--117,
  [\href{http://arxiv.org/abs/1212.3620}{{\tt arXiv:1212.3620}}].

\bibitem{Alwall:2007fs}
J.~Alwall et~al., {\it {Comparative study of various algorithms for the merging
  of parton showers and matrix elements in hadronic collisions}},  {\em Eur.
  Phys. J.} {\bf C53} (2008) 473--500,
  [\href{http://arxiv.org/abs/0706.2569}{{\tt arXiv:0706.2569}}].

\bibitem{Aad:2014fha}
{\bf ATLAS} Collaboration, G.~Aad et~al., {\it {Search for new resonances in
  $W\gamma$ and $Z\gamma$ final states in $pp$ collisions at $\sqrt s=8$ TeV
  with the ATLAS detector}},  {\em Phys. Lett.} {\bf B738} (2014) 428--447,
  [\href{http://arxiv.org/abs/1407.8150}{{\tt arXiv:1407.8150}}].

\bibitem{ATLAS-CONF-2016-010}
{\it {Search for heavy resonances decaying to a $Z$ boson and a photon in $pp$
  collisions at $\sqrt{s}=13$ TeV with the ATLAS detector}},  Tech. Rep.
  ATLAS-CONF-2016-010, CERN, Geneva, Mar, 2016.

\bibitem{Aad:2015kna}
{\bf ATLAS} Collaboration, G.~Aad et~al., {\it {Search for an additional, heavy
  Higgs boson in the $H\rightarrow ZZ$ decay channel at $\sqrt{s} = 8\;\text{
  TeV }$ in $pp$ collision data with the ATLAS detector}},  {\em Eur. Phys. J.}
  {\bf C76} (2016), no.~1 45, [\href{http://arxiv.org/abs/1507.05930}{{\tt
  arXiv:1507.05930}}].

\bibitem{ATLAS-CONF-2015-071}
{\it {Search for diboson resonances in the llqq final state in pp collisions at
  $\sqrt{s}$ = 13 TeV with the ATLAS detector}},  Tech. Rep.
  ATLAS-CONF-2015-071, CERN, Geneva, Dec, 2015.

\bibitem{Aad:2015agg}
{\bf ATLAS} Collaboration, G.~Aad et~al., {\it {Search for a high-mass Higgs
  boson decaying to a $W$ boson pair in $pp$ collisions at $\sqrt{s} = 8$ TeV
  with the ATLAS detector}},  {\em JHEP} {\bf 01} (2016) 032,
  [\href{http://arxiv.org/abs/1509.00389}{{\tt arXiv:1509.00389}}].

\bibitem{ATLAS-CONF-2016-021}
{\it {Search for a high-mass Higgs boson decaying to a pair of W bosons in pp
  collisions at sqrt(s)=13 TeV with the ATLAS detector}},  Tech. Rep.
  ATLAS-CONF-2016-021, CERN, Geneva, Apr, 2016.

\bibitem{Chatrchyan:2013lca}
{\bf CMS} Collaboration, S.~Chatrchyan et~al., {\it {Searches for new physics
  using the $t\bar{t}$ invariant mass distribution in pp collisions at
  $\sqrt{s}$=8  TeV}},  {\em Phys. Rev. Lett.} {\bf 111} (2013), no.~21
  211804, [\href{http://arxiv.org/abs/1309.2030}{{\tt arXiv:1309.2030}}].
  [Erratum: Phys. Rev. Lett.112,no.11,119903(2014)].

\bibitem{ATLAS-CONF-2016-014}
{\it {Search for heavy particles decaying to pairs of highly-boosted top quarks
  using lepton-plus-jets events in proton--proton collisions at $\sqrt{s} = 13$
  TeV with the ATLAS detector}},  Tech. Rep. ATLAS-CONF-2016-014, CERN, Geneva,
  Mar, 2016.

\bibitem{Aad:2015zva}
{\bf ATLAS} Collaboration, G.~Aad et~al., {\it {Search for new phenomena in
  final states with an energetic jet and large missing transverse momentum in
  pp collisions at $\sqrt{s}=$8 TeV with the ATLAS detector}},  {\em Eur. Phys.
  J.} {\bf C75} (2015), no.~7 299, [\href{http://arxiv.org/abs/1502.01518}{{\tt
  arXiv:1502.01518}}]. [Erratum: Eur. Phys. J.C75,no.9,408(2015)].

\bibitem{Aaboud:2016tnv}
{\bf ATLAS} Collaboration, M.~Aaboud et~al., {\it {Search for new phenomena in
  final states with an energetic jet and large missing transverse momentum in
  $pp$ collisions at $\sqrt{s}=13$ TeV using the ATLAS detector}},
  \href{http://arxiv.org/abs/1604.07773}{{\tt arXiv:1604.07773}}.

\bibitem{Abdesselam:2010pt}
A.~Abdesselam et~al., {\it {Boosted objects: A Probe of beyond the Standard
  Model physics}},  {\em Eur. Phys. J.} {\bf C71} (2011) 1661,
  [\href{http://arxiv.org/abs/1012.5412}{{\tt arXiv:1012.5412}}].

\bibitem{Altheimer:2012mn}
A.~Altheimer et~al., {\it {Jet Substructure at the Tevatron and LHC: New
  results, new tools, new benchmarks}},  {\em J. Phys.} {\bf G39} (2012)
  063001, [\href{http://arxiv.org/abs/1201.0008}{{\tt arXiv:1201.0008}}].

\bibitem{ATLAS-CONF-2015-068}
{\it {Search for diboson resonances in the $\nu\nu qq$ final state in $pp$
  collisions at $\sqrt{s}=$13 TeV with the ATLAS detector}},  Tech. Rep.
  ATLAS-CONF-2015-068, CERN, Geneva, Dec, 2015.

\bibitem{Alloul:2013bka}
A.~Alloul, N.~D. Christensen, C.~Degrande, C.~Duhr, and B.~Fuks, {\it
  {FeynRules 2.0 - A complete toolbox for tree-level phenomenology}},  {\em
  Comput. Phys. Commun.} {\bf 185} (2014) 2250--2300,
  [\href{http://arxiv.org/abs/1310.1921}{{\tt arXiv:1310.1921}}].

\bibitem{Alwall:2014hca}
J.~Alwall, R.~Frederix, S.~Frixione, V.~Hirschi, F.~Maltoni, O.~Mattelaer,
  H.~S. Shao, T.~Stelzer, P.~Torrielli, and M.~Zaro, {\it {The automated
  computation of tree-level and next-to-leading order differential cross
  sections, and their matching to parton shower simulations}},  {\em JHEP} {\bf
  07} (2014) 079, [\href{http://arxiv.org/abs/1405.0301}{{\tt
  arXiv:1405.0301}}].

\bibitem{Ball:2012cx}
R.~D. Ball et~al., {\it {Parton distributions with LHC data}},  {\em Nucl.
  Phys.} {\bf B867} (2013) 244--289,
  [\href{http://arxiv.org/abs/1207.1303}{{\tt arXiv:1207.1303}}].

\bibitem{Knapen:2015dap}
S.~Knapen, T.~Melia, M.~Papucci, and K.~Zurek, {\it {Rays of light from the
  LHC}},  {\em Phys. Rev.} {\bf D93} (2016), no.~7 075020,
  [\href{http://arxiv.org/abs/1512.04928}{{\tt arXiv:1512.04928}}].

\bibitem{Bellazzini:2015nxw}
B.~Bellazzini, R.~Franceschini, F.~Sala, and J.~Serra, {\it {Goldstones in
  Diphotons}},  {\em JHEP} {\bf 04} (2016) 072,
  [\href{http://arxiv.org/abs/1512.05330}{{\tt arXiv:1512.05330}}].

\bibitem{Sato:2016hls}
R.~Sato and K.~Tobioka, {\it {LHC Future Prospects of the 750 GeV Resonance}},
  \href{http://arxiv.org/abs/1605.05366}{{\tt arXiv:1605.05366}}.

\bibitem{No:2016htu}
J.~M. No, {\it {Is it $SU(2)_{\mathrm{L}}$ or just $U(1)_{\mathrm{Y}}$? $750$
  GeV di-photon probes of the electroweak nature of new states}},
  \href{http://arxiv.org/abs/1605.05900}{{\tt arXiv:1605.05900}}.

\bibitem{Thamm:2015zwa}
A.~Thamm, R.~Torre, and A.~Wulzer, {\it {Future tests of Higgs compositeness:
  direct vs indirect}},  {\em JHEP} {\bf 07} (2015) 100,
  [\href{http://arxiv.org/abs/1502.01701}{{\tt arXiv:1502.01701}}].

\bibitem{Buttazzo:2015bka}
D.~Buttazzo, F.~Sala, and A.~Tesi, {\it {Singlet-like Higgs bosons at present
  and future colliders}},  {\em JHEP} {\bf 11} (2015) 158,
  [\href{http://arxiv.org/abs/1505.05488}{{\tt arXiv:1505.05488}}].

\bibitem{jsterling}
J.~W. Sterling. personal communication.

\bibitem{Aad:2015wra}
{\bf ATLAS} Collaboration, G.~Aad et~al., {\it {Search for a CP-odd Higgs boson
  decaying to Zh in pp collisions at $\sqrt{s} = 8$ TeV with the ATLAS
  detector}},  {\em Phys. Lett.} {\bf B744} (2015) 163--183,
  [\href{http://arxiv.org/abs/1502.04478}{{\tt arXiv:1502.04478}}].

\bibitem{ATLAS-CONF-2016-015}
{\it {Search for a CP-odd Higgs boson decaying to Zh in pp collisions at √s =
  13 TeV with the ATLAS detector}},  Tech. Rep. ATLAS-CONF-2016-015, CERN,
  Geneva, Mar, 2016.

\bibitem{Aad:2014cka}
{\bf ATLAS} Collaboration, G.~Aad et~al., {\it {Search for high-mass dilepton
  resonances in pp collisions at $\sqrt{s}=8$??TeV with the ATLAS detector}},
  {\em Phys. Rev.} {\bf D90} (2014), no.~5 052005,
  [\href{http://arxiv.org/abs/1405.4123}{{\tt arXiv:1405.4123}}].

\bibitem{ATLAS-CONF-2015-070}
{\it {Search for new phenomena in the dilepton final state using proton-proton
  collisions at √ s = 13 TeV with the ATLAS detector}},  Tech. Rep.
  ATLAS-CONF-2015-070, CERN, Geneva, Dec, 2015.

\bibitem{Aad:2014xka}
{\bf ATLAS} Collaboration, G.~Aad et~al., {\it {Search for resonant diboson
  production in the $\mathrm {\ell \ell }q\bar{q}$ final state in $pp$
  collisions at $\sqrt{s} = 8$ TeV with the ATLAS detector}},  {\em Eur. Phys.
  J.} {\bf C75} (2015) 69, [\href{http://arxiv.org/abs/1409.6190}{{\tt
  arXiv:1409.6190}}].

\bibitem{Aad:2015uka}
{\bf ATLAS} Collaboration, G.~Aad et~al., {\it {Search for Higgs boson pair
  production in the $b\bar{b}b\bar{b}$ final state from pp collisions at
  $\sqrt{s} = 8$ TeVwith the ATLAS detector}},  {\em Eur. Phys. J.} {\bf C75}
  (2015), no.~9 412, [\href{http://arxiv.org/abs/1506.00285}{{\tt
  arXiv:1506.00285}}].

\bibitem{ATLAS-CONF-2016-017}
{\it {Search for pair production of Higgs bosons in the $b\bar{b}b\bar{b}$
  final state using proton-proton collisions at $\sqrt{s} = 13$ TeV with the
  ATLAS detector}},  Tech. Rep. ATLAS-CONF-2016-017, CERN, Geneva, Mar, 2016.

\bibitem{Aad:2015mna}
{\bf ATLAS} Collaboration, G.~Aad et~al., {\it {Search for high-mass diphoton
  resonances in $pp$ collisions at $\sqrt{s}=8$ TeV with the ATLAS detector}},
  {\em Phys. Rev.} {\bf D92} (2015), no.~3 032004,
  [\href{http://arxiv.org/abs/1504.05511}{{\tt arXiv:1504.05511}}].

\end{thebibliography}\endgroup

\end{document}